\pdfoutput=1
\RequirePackage{xcolor}
\documentclass[journal,twoside,web]{ieeecolor}
\usepackage{tmi}
\usepackage{cite}
\usepackage{amsmath,amssymb,amsfonts}
\usepackage{algorithmic}
\usepackage{graphicx}
\usepackage{float}
\usepackage{dblfloatfix}
\usepackage{textcomp}
\usepackage{booktabs} 
\usepackage{multirow} 
\usepackage{subcaption} 
\usepackage[colorlinks=true, urlcolor=blue, linkcolor=black]{hyperref}
\usepackage{tabularx,tabulary}

\usepackage{array}
\usepackage{siunitx}
\usepackage{scalerel}
\usepackage{tikz}
\usetikzlibrary{svg.path}

\definecolor{orcidlogocol}{HTML}{A6CE39}
\tikzset{
	orcidlogo/.pic={
		\fill[orcidlogocol] svg{M256,128c0,70.7-57.3,128-128,128C57.3,256,0,198.7,0,128C0,57.3,57.3,0,128,0C198.7,0,256,57.3,256,128z};
		\fill[white] svg{M86.3,186.2H70.9V79.1h15.4v48.4V186.2z}
		svg{M108.9,79.1h41.6c39.6,0,57,28.3,57,53.6c0,27.5-21.5,53.6-56.8,53.6h-41.8V79.1z M124.3,172.4h24.5c34.9,0,42.9-26.5,42.9-39.7c0-21.5-13.7-39.7-43.7-39.7h-23.7V172.4z}
		svg{M88.7,56.8c0,5.5-4.5,10.1-10.1,10.1c-5.6,0-10.1-4.6-10.1-10.1c0-5.6,4.5-10.1,10.1-10.1C84.2,46.7,88.7,51.3,88.7,56.8z};
	}
}

\newcommand\orcidicon[1]{\href{https://orcid.org/#1}{\mbox{\scalerel*{
				\begin{tikzpicture}[yscale=-1,transform shape]
				\pic{orcidlogo};
				\end{tikzpicture}
			}{|}}}}
\usepackage{cleveref}
\crefname{subfigure}{figs.}{Figs.}

\usepackage{tikz,libertine}
\usetikzlibrary{calc,spy}
\tikzstyle{closeup} = [
opacity=1.0,          
height=0.9cm,         
width=0.9cm,          
connect spies, mblue  
]
\tikzstyle{largewindow} = [mblue, line width=0.50mm]
\tikzstyle{smallwindow} = [mblue, line width=0.20mm]
\usepackage{tikz}
\usetikzlibrary{quotes,arrows.meta,positioning,3d, calc}
\usepackage{xcolor}
\definecolor{NavyBlue}{rgb}{0,0,0.5}

\def\BibTeX{{\rm B\kern-.05em{\sc i\kern-.025em b}\kern-.08em
    T\kern-.1667em\lower.7ex\hbox{E}\kern-.125emX}}
\begin{document}
\title{
RF-ULM: Ultrasound Localization Microscopy Learned from Radio-Frequency Wavefronts
}

\author{Christopher Hahne$^{\star}$$^{\textsuperscript{\orcidicon{0000-0003-2786-9905}}}$, 
Georges Chabouh$^{\textsuperscript{\orcidicon{0000-0003-0760-909X}}}$, 
Arthur Chavignon$^{\textsuperscript{\orcidicon{0000-0001-7883-7482}}}$, 
Olivier Couture$^{\textsuperscript{\orcidicon{0000-0002-8683-1424}}}$,
and
Raphael Sznitman$^{\textsuperscript{\orcidicon{0000-0001-6791-4753}}}$
\thanks{This work was supported in part by the Hasler Foundation under Grant number 22027. $^{\star}$Corresponding author email: \href{mailto:christopher.hahne@unibe.ch}{\textcolor{nblue}{christopher.hahne [ät] unibe.ch}}
}
\thanks{C. Hahne and R. Sznitman are with the 
Artificial Intelligence in Medical Imaging Laboratory, ARTORG Center, University of Bern,
Bern, Switzerland.
}
\thanks{G. Chabouh, A. Chavignon and O. Couture are with the Laboratoire d'Imagerie Biomédicale,
Inserm, CNRS, Sorbonne Université,
Paris, France.}
}
\newcommand{\CH}[1]{{\color{blue}{\bf CH: #1}}}
  \newcommand{\ch}[1]{{\color{blue}{#1}}}
\maketitle

\begin{abstract} 
In Ultrasound Localization Microscopy~\mbox{(ULM)}, achieving high-resolution images relies on the precise localization of contrast agent particles across a series of beamformed frames. %
However, our study uncovers an enormous potential: The process of delay-and-sum beamforming leads to an irreversible reduction of Radio-Frequency~\mbox{(RF)} channel data, while its implications for localization remain largely unexplored. 
The rich contextual information embedded within RF wavefronts, including their hyperbolic shape and phase, offers great promise for guiding Deep Neural Networks~\mbox(DNNs) in challenging localization scenarios. To fully exploit this data, we propose to directly localize scatterers in RF channel data.
Our approach involves a custom super-resolution DNN using learned feature channel shuffling, non-maximum suppression, and a semi-global convolutional block for reliable and accurate wavefront localization. Additionally, we introduce a geometric point transformation that facilitates seamless mapping to the B-mode coordinate space. To understand the impact of beamforming on ULM, we validate the effectiveness of our method by conducting an extensive comparison with State-Of-The-Art (SOTA) techniques.
%
We present the inaugural \textit{in vivo} results from a wavefront-localizing DNN, highlighting its real-world practicality. Our findings show that RF-ULM bridges the domain shift between synthetic and real datasets, 
offering a considerable advantage in terms of precision and complexity. 
To enable the broader research community to benefit from our findings, our code and the associated SOTA methods are made available at \texttt{{\color{mblue}\textmd{\url{{https://github.com/hahnec/rf-ulm}}}}}.
\end{abstract}


\section{Introduction}
\label{sec:introduction}
\begin{figure}[!t]
    \centering
    \begin{minipage}[b]{\linewidth}
        \centerline{\resizebox{\linewidth}{!}{\input{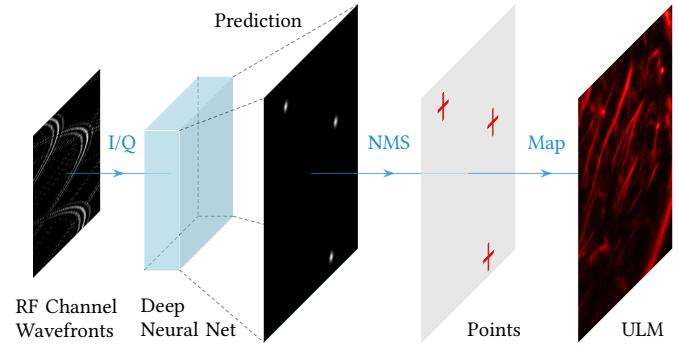}}}
    \end{minipage}
    \caption{\textbf{Overview of the RF-ULM framework:} We leverage RF channel data by feeding In-phase and Quadrature (I/Q) components into a super-resolution neural network. This enables microbubble localization through Non-Maximum Suppression (NMS) without relying on Delay-And-Sum (DAS) beamforming. The resulting sub-wavelength localizations are then mapped to the \mbox{B-mode} coordinate space using an affine transformation. The final ULM rendering step involves the accumulation of all detections over time
    .}
    \label{fig:overview}
\end{figure}

%
\IEEEPARstart{I}{n} the realm of Ultrasound Localization Microscopy (ULM), a compelling opportunity emerges: unlocking the hidden potential of Radio-Frequency (RF) channel data for precise particle localization that liberates ULM from the constraints of conventional beamforming methods. 
A schematic overview of our approach is outlined in Fig.~\ref{fig:overview}.

Contrast-Enhanced-UltraSound (CEUS) suffers from low image resolution due to the diffraction limit, making it less effective compared to modalities like microangio-computed tomography~\cite{heiles2022pala}.
In response to this constraint, ULM emerged as a transformative approach~\cite{errico2015ultrafast,StateOfTheArt,song:2018,van2020super,liu2020deep,gulm:2023} surpassing the diffraction limit. This is achieved by pinpointing contrast agent particles, commonly referred to as MicroBubbles (MBs)~\cite{chabouh2021spherical,chabouh:2023:buckling}, across a series of frames. Accurate and reliable localization of MBs thus became a central research topic in recent years. 




Since Errico~\textit{et al.}\cite{errico2015ultrafast} pioneered the high-resolution ULM imaging capability, there has been a notable surge in research interest
~\cite{StateOfTheArt,song:2018,van2020super,liu2020deep,gulm:2023}. This work has not only advanced our understanding of ULM's technical capabilities but has also laid the foundation for its clinical adoption. Recent research endeavors have focused on in-human ULM applications, including breast lesion characterization~\cite{bar:2021}, brain vascularization~\cite{demene2021transcranial} and kidney blood flow assessment~\cite{bodard2023ultrasound,song2023super}.
ULM's capabilities have been extended to \mbox{encompass three-dimensional} (3-D) models~\cite{chavignon:2021,heiles2022volumetric}, allowing for the visualization of coronary vascular flow~\cite{demeulenaere2022coronary} or different types of stroke~\cite{chavignon20223d}.

Several localization algorithms have been hand-crafted for ULM, including deconvolution~\cite{errico2015ultrafast}, two-dimensional \mbox{(2-D)} fitting~\cite{song:2018}, or Radial Symmetry~\mbox{(RS)}~\cite{heiles2022pala}. To estimate blood flow velocities and handle false positive detections, researchers introduced localization tracking algorithms. This can be realized using the Munkres linker~\cite{heiles2022pala} or a multi-feature Kalman approach~\cite{yan2022super}. As an alternative to tracking, the Curvelet Transform-based Sparsity Promoting (CTSP)~\cite{you2022curvelet} has shown to help recover MB positions from short acquisitions.

With the rise of deep learning in the past decade, several deep learning frameworks have been adopted for ULM to overcome low MB perfusion and enhance detection reliability from fewer frames\cite{van2020super,liu2020deep,hahne:23:ius}. Specifically, the ULTRA-SR challenge has catalyzed a diverse array of studies that leverage cutting-edge deep learning architectures, including U-Nets~\cite{wang2022general,long2022super}, Generative Adversarial Networks (GANs)\cite{sui2022generative}, and Transformers\cite{gharamaleki2022transformer,liu2022ultrasound}. 
Recently,
learning-based ULM research has taken the direction of temporal-aware localization. This is achieved by the integration of temporal data modules into Deep Neural Networks (DNNs)~\cite{milecki2021deep,chen2022deep,chen2023deep}. 


Notably, ultrasound-based image formation has recently been accomplished without beamforming~\cite{nair2018deep,zhang2021ultrasound,gulm:2023}. This paradigm shift is exemplified by DNNs, which exhibit significant promise in reconstructing objects in absence of Delay-And-Sum~\mbox{(DAS)} beamforming~\cite{nair2018deep,zhang2021ultrasound}. RF-based localization has also been tackled without DNNs, for example, using wavefront shape regression~\cite{couture2011microbubble}. As an alternative, \mbox{Geometric-ULM} \mbox{(G-ULM)} recently achieved image recovery through trilateration using Time-of-Arrival detections to pinpoint MBs in \mbox{B-mode} coordinate space~\cite{gulm:2023}. While previous work examined the impact of beamformers on localization~\cite{corazza2022beamforming}, RF-trained networks recently garnered increasing attention~\cite{youn2020detection,blanken2022sr} due to their capability to skip beamforming in DNN-based ULM pipelines~\cite{van2020super,liu2020deep}. This evolving landscape underscores the growing significance of RF-based localization and its potential for ULM rendering.

Nonetheless, previous work in RF-based ULM has several noteworthy limitations: Existing attempts show image reconstruction using phantom data lacking \textit{in vivo} comparison~\cite{youn2020detection,blanken2022sr}, which raises the question on the practicability and performance. In particular, spatio-temporal filtering is a common practice for beamformed ULM, yet its impact on RF input has been overlooked. %
Moreover, the computational complexity associated with G-ULM~\cite{gulm:2023} poses a challenge for its real-world application. Also, prior works neglected to address the fusion of localizations from compounded waves. 
This gap in knowledge prompts crucial inquiries on the benefits of RF-based ULM for \textit{in vivo} scenarios. %
\par
To this end, we propose a novel framework to advance ULM image rendering through a fast \textit{in vivo} RF-based localization at sub-wavelength precision. %
Until now, the prevailing approach has been to utilize beamformed images as the primary input for the localization.
However, we examine this notion by exploring the hypothesis that beamforming, as a hand-crafted focusing method, may not be the most efficient localization step. %
The summation in beamforming reduces wavefront information irretrievably, which becomes evident when attempting to reverse the process: %
While RF wavefronts can be transformed into \mbox{B-mode} images, recovering the original wavefront signal from a beamformed image is an ill-posed inverse problem. This observation suggests that raw channel data generally contains the utmost information %
offering the potential for most accurate scatterer localization when properly analyzed. This motivates us to bypass beamforming and enhance ULM by letting a network learn RF properties.
Thereby, we also address the pressing computational demands arising from the need to beamform thousands of images. 
Since we train localization exclusively on \textit{in silico} data, the wavefront information plays a crucial role in enhancing the generalization capabilities of the network to unseen \textit{in vivo} inputs, known as the domain gap.
%
In contrast to existing studies, we feed RF data into our customized super-resolution Deep Neural Network~\mbox{(DNN)} to obtain distinct scatterer positions from fast Non-Maximum Suppression~\mbox{(NMS)} and a geometric transformation, as illustrated in Fig.~\ref{fig:overview}. This concept can be envisioned as swapping a conventional beamformer for an efficient super-resolution DNN. Combined with an effortless point extraction, our novel and cost-effective geometric mapping between \mbox{B-mode} and RF coordinates enables ULM rendering and training from RF data while mitigating the computational complexity as imposed by G-ULM~\cite{gulm:2023}. %
To harness these advancements, we conduct a benchmark analysis with state-of-the-art methods, including an ablation study to evaluate our framework tailored for RF signals. %
Our proposed DNN pipeline outperforms state-of-the-art methods in terms of localization accuracy while achieving competitive processing times. For reproducibility, we release our original code as well as the state-of-the-art implementations. 
\par
In this study, the primary focus is to independently evaluate the standalone localization performance of contemporary networks.
Tracking, which helps overcome low frame rates and measure velocities, is not incorporated into our main evaluation since it is considered a post-processing step. 
\par
While this introduction provides an in-depth review of the relevant literature and our motivation, subsequent sections of this paper delve into a comprehensive analysis of RF-based localization. In Section~\ref{sec:theory}, we present the methodology and data sources employed in our study, elaborating the rationale behind our chosen approach. The empirical findings and results are provided in Section~\ref{sec:results}, where we offer a thorough examination of ULM rendering from \mbox{B-mode} and RF data. 
In Section~\ref{sec:conclusion}, we consolidate our findings and suggest avenues for future research, 
culminating in a holistic understanding of ULM in the absence of computational beamforming. 
\section{Theory}
\label{sec:theory}
ULM can be framed as a localization problem within the two-dimensional \mbox{(2-D)} signal containing spatio-temporal wavefronts. Image-based localization is a well-studied task in the computer vision field such that deterministic algorithms have been explored for ULM in previous research~\cite{song:2018,heiles2022pala}. While recent developments in super-resolution DNNs have gained attention in the ULM community~\cite{van2020super,liu2020deep}, we wish to extend on this work by investigating the impact of beamforming and the potential of RF channel data. 

\begin{figure*}[t!]
    \begin{minipage}[b]{\linewidth}
        \centerline{\resizebox{\linewidth}{!}{\input{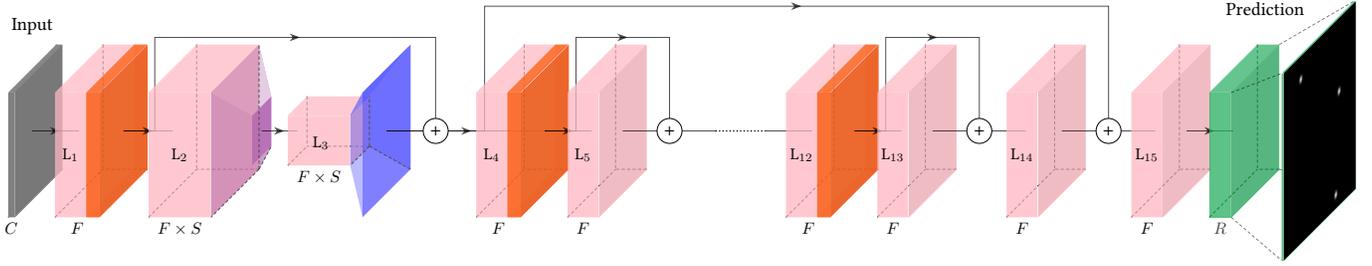}}}
    \end{minipage}
    \caption{\textbf{Our Semi-Global-SPCN architecture} employs multiple convolutional layers, residual skip connections and channel shuffling to predict a map upsampled by factor $R$. The model takes as input a \mbox{2-D} signal with $C$ channels for optional feature concatenation. The initial layer applies a \mbox{2-D} convolution (opaque pink) with $F$ filters and a kernel size of 9 followed by a Rectified Linear Unit (ReLU) (dark orange). Layer 2 and 3 represent our proposed semi-global bottleneck block consisting of \mbox{2-D} convolutions with a kernel size of 5, $S=\max(1,G/10)$, LeakyReLUs, and down- as well as upsampling blocks (purple and blue) with scale $G=16$, respectively. The subsequent layers (4 to 14) consist of \mbox{2-D} convolutions with $F$ filters and a kernel size of 7. Residual connections are added after every other layer, whereas a ReLU follows convolutions without residual connections. The second last layer uses a \mbox{2-D} convolution with $F$ filters and a kernel size of 3, followed by an element-wise addition with the third layer residual output. The final output is obtained by applying a \mbox{2-D} convolution with the specified upsampling factor $R$ and a kernel size of 3, followed by a channel to pixel shuffle operation (green).}\label{fig:arch}
\end{figure*}

\subsection{Semi-Global-SPCN Architecture}

The rapidly advancing field of deep learning has introduced a multitude of architectures suitable for tackling this localization challenge~\cite{ronneberger2015u,shi2016real,lim2017enhanced}. Notably, many of these models adopt a design paradigm characterized by a contracting and expanding path, reminiscent of the U-Net architecture~\cite{ronneberger2015u}. This trend is also observable in the domain of ultrasound imaging~\cite{nair2018deep,li2019cr,zhang2021ultrasound}, particularly in the context of ULM~\cite{van2020super,wang2022general,long2022super}.

The rationale behind employing the U-Net structure lies in its ability to establish global context for image objects that are larger than the convolution kernel size. The contraction aids in correlating pixels spaced farther apart to form cohesive global segments, fostering a comprehensive understanding of the underlying signal characteristics.

While this proves useful in the context of semantic segmentation, sample reduction carries the risk of sacrificing the ability to recover essential signal details and may be redundant for sub-wavelength localization. Rather than employing a bottleneck contraction, recent advancements in efficient image super-resolution networks have chosen to forgo spatial downscaling. Instead, they expand the feature channels and incorporate upsampling through a trailing feature channel shuffle operation, as outlined in previous works~\cite{shi2016real,lim2017enhanced}. For example, Liu \textit{et al.} proposed a modified Sub-Pixel Convolutional Network (mSPCN)~\cite{liu2020deep} as a flavored \mbox{super-resolution} DNN for ULM. However, such networks tend to fall short in capturing extensive contextual information, which becomes crucial when dealing with larger signal regions such as RF wavefronts. In this case, a signal contraction may facilitate a network to recognize and pinpoint the tip of spatially extended wavefront signals. 

For these reasons, we present a customized network architecture inspired by previous super-resolution networks~\cite{shi2016real,lim2017enhanced} to address the challenge of \mbox{2-D} RF wavefront localization. Our approach aims to balance between the previously mentioned requirements, with a focus on highly accurate and reliable wavefront detection. 
Unlike a U-Net model~\cite{van2020super}, our network does not include a global bottleneck contraction. This omission is deliberate, as it allows us to preserve essential resolution information and save on computational complexity. Instead, we address localization refinement across large contextual regions by introducing a unique element into our \mbox{Semi-Global} \mbox{Sub-Pixel} Convolutional Network \mbox{(SG-SPCN)} architecture: a solitary bottleneck block for \mbox{semi-global} context recognition, placed at an early stage of the network. A visual representation of our network is depicted in Fig.~\ref{fig:arch}. Our \mbox{SG-SPCN} model is designed to handle variable input lengths, and its block dimensions are parameterizable. However, we adhere to the conventions of the image super-resolution field, setting the number of feature channels $F$ to 64. Also, we vary the upsampling factor with $R=12$ for real data and $R=8$ using synthetic data for fair comparison with concurrent networks.
The ideal scale depends on the transducer arrangement and spatial extent of arriving wavefronts. We heuristically approximate the width of distant wavefronts by measuring the spacing between samples where the maximum amplitude has dropped by $50\%$. In our \textit{insilico} PALA dataset, this amounts to about 65 samples width. To accomplish a receptive field $r_f$ of sample size 65, we then calculate the semi-global scale by $G=(r_f-1)/(k_1-1)=(65-1)/(5-1)=16$ where $k_1=5$ is the convolution kernel size.

To assess the efficacy of our proposed architecture, we conduct an ablation analysis in \ref{sec:ablation}. Our evaluation of \mbox{RF-ULM} includes a comparison with unsupervised ULM techniques~\cite{heiles2022pala,gulm:2023}, and established \mbox{2-D} adaptations of models commonly employed in computer vision~\cite{shi2016real,lim2017enhanced}.

\subsection{Training, Augmentation and Inference}
\textbf{Training:} 
%
We learn neural network weights akin to~\cite{van2020super,liu2020deep} with modifications detailed hereafter. Our approach bypasses beamforming and directly feeds RF channel data following In-phase and Quadrature~(I/Q) demodulation. 
We construct a model input $\mathbf{X}$ by stacking the complex I/Q components as feature channels. This ensures more efficient retention of crucial information than in highly sampled RF channels or magnitude B-mode frames.

The training loss function $\mathcal{L}(\cdot)$ is defined as follows:
\begin{align}
\mathcal{L}(\mathbf{X},\mathbf{Y})= \lVert f(\mathbf{X}) - \lambda_0(\mathbf{G}_{\sigma} \circledast \mathbf{Y})\rVert_2^2 + \lambda_1 \lVert f(\mathbf{X})\rVert_1
\label{eq:loss}
\end{align}

Here, $\mathbf{G}_{\sigma}$ denotes a 2-D Gaussian kernel used for convolution $\circledast$ with label map $\mathbf{Y}$ amplified by $\lambda_0$. %
We use a constant $\sigma=1$ kernel width~\cite{van2020super,liu2020deep} except for \mbox{$R>10$} where we gradually decrease $\sigma$ after each epoch. The intuition is to facilitate steep loss improvements early on for frames where the background dominates over segments of interest (MB pixels). We employ an inverse quadratic decrease, varying $\sigma$ from 3.5 to 1, which refines the spatial extent of localizations as the training progresses. %
%
The second loss term is an $L_1$ regularization term, scaled by $\lambda_1$, which prevents $f(\cdot)$ from predicting an excessive number of false positives.
%
We train with an Adam optimizer for a maximum of 40 epochs, employing a batch size of 16, weight decay set at $1\mathrm{e}{-8}$, and an initial learning rate of $1\mathrm{e}{-3}$. The learning rate schedule is implemented using cosine annealing. For regularization, the scaling factors are chosen as follows: $\lambda_0=(\max(\mathbf{G}_{\sigma}\circledast\mathbf{Y})/120)^{-1}$ and $\lambda_1=1\mathrm{e}{-2}$. 

\begin{table}[h!]
    \centering
    \caption{Notation symbols and definitions}
    \begin{tabular}{|p{0.407\linewidth}|p{0.483\linewidth}|}
        \hline
        \textbf{Symbol} & \textbf{Definition} \\
        \hline
        $R$ & Spatial upscale factor \\
        \hline
        $\mathbf{X} \in \left[-1,1\right]^{2 \times U \times V}$ & I/Q channel frame with shape $U,V$ \\
        \hline
        $\mathbf{Y} \in \{0,1\}^{RU \times RV}$ & Scatterer label with shape $U,V$ \\
        \hline
        $\mathbf{G}_{\sigma} \in [0,1]^{(7+R) \times (7+R)}$ & 2-D Gaussian kernel with scale $\sigma$ \\
        \hline
        $f(\cdot):\mathbb{R}^{2\times U\times V}\mapsto\mathbb{R}^{RU\times RV}$ & SG-SPCN as a function \\
        \hline
        $\lambda_0$, $\lambda_1$ & Label scale, $L_1$ regularization scale\\
        \hline
        $\mathbf{v}_s, \mathbf{x}_k \in \mathbb{R}^3$ & Virtual source, transducer positions \\
        \hline
        $c_s, f_s$ & Speed of sound, sample rate \\
        \hline
        $\mathbf{p}_i \in \mathbb{R}^3$ & GT point at index $i$ in B-mode space \\
        \hline
        $\mathbf{p}_{i,k}' \in \mathbb{R}^{3\times K}$ & Point projections at $K$ transducers \\
        \hline
        $\mathbf{p}_i^\star \in \mathbb{R}^3$ & GT wavefront point in channel space \\
        \hline
        $\mathbf{A} \in \mathbb{R}^{2 \times 3}$ & Affine point transformation matrix \\
        \hline
    \end{tabular}
\end{table}

\textbf{Augmentation:} 
Data augmentation plays a pivotal role in enhancing the robustness and generalization of DNNs. To address the challenges posed by variations in the input data, we employ 
random frame cropping, random flips along the axis orthogonal to the transducer, occasional Gaussian blurring, and random rotation within 5 degrees angle. We also add clutter noise according to~\cite{heiles2022pala} during training with a signal-to-noise ratio of $50~\text{dB}$ for each frame to enhance its robustness against noise. In addition, we normalize amplitudes ensuring that inputs are $\mathbf{X}\in\left[-1,1\right]^{2\times U\times V}$. These augmentations mitigate overfitting and enhance the model's ability to learn from diverse signal patterns to better handle unseen data. 

\textbf{Inference:} A map predicted by $f(\mathbf{X})$ provides localization probabilities \if at each coordinate \fi in an equidistant sampling grid. Each capture of one transmit event is processed individually in a parallelized fashion. %
To pinpoint scatterer coordinates from DNN predictions, we introduce NMS-based thresholding. Initially, $f(\mathbf{X})$ undergoes NMS~\cite{neubeck2006nms}, implemented via \textit{max-pooling} and \textit{fancy indexing}, which is complemented by thresholding to filter out localizations with low probability. This threshold is estimated by geometric-mean analysis of the Receiver Operating Characteristic (ROC) curve~\cite{kubat1997addressing}. %
While strictly applied to \textit{in silico} data, we heuristically increase the ROC-based threshold for the \textit{in vivo} domain to mitigate false positives. %
%
Given the NMS-based maxima at upsampled integer coordinates, we rescale localizations to the sub-pixel precise input resolution. These positions represent points in transducer channel space, which have to be transferred to \mbox{B-mode} points for comparison. 

\subsection{Coordinate Space Transformation}
Leveraging a wavefront localization framework requires a coordinate conversion from channel data to \mbox{B-mode} space and vice versa. To alleviate the computational complexity associated with G-ULM~\cite{gulm:2023}, we map points between \mbox{B-mode} and channel data coordinate space using an affine transformation algebra that is explained hereafter.

\textbf{Forward Label Projection:} Since scientists discovered ULM, Ground Truth (GT) labels are generally provided in \mbox{B-mode} coordinate space. However, learning localization directly from transducer channels requires to map these labels to the channel coordinate space. Following the transducer geometry proposed in G-ULM, we project GT points to the channel domain based on the Time-of-Flight (ToF) physics.

Let GT point labels be given by $\mathbf{p}_i=\left[y_i,z_i,1\right]^\intercal$ with index $i$ in \mbox{B-mode} space. The points $y_i$ and $z_i$ represent lateral and axial coordinates, which we project to the channels using, 
\begin{align}
      %
      \mathbf{p}_{i,k}'= \frac{f_s}{c_s}\Big(\lVert\mathbf{p}_i-\mathbf{v}_s\rVert_2+\lVert\mathbf{p}_i-\mathbf{x}_k\rVert_2-s\Big)  
      \, , \quad \forall k \, ,
      \label{eq:project}
\end{align}
where $\mathbf{v}_s\in\mathbb{R}^3$ is the virtual transducer source, $\mathbf{x}_k\in\mathbb{R}^3$ is a transducer position with index $k\in\{1,2,\dots,K\}$ and $\lVert\cdot\rVert_2$ is the Euclidean norm. Here, $s$ deducts the travel distance for the elapsed time between emission and capture start. The scalar $c_s$ denotes the speed of sound and $f_s$ the sample rate.

Equation \eqref{eq:project} demonstrates that a single \mbox{B-mode} point $\mathbf{p}_i$ yields one label $\mathbf{p}'_{i,k}$ per channel $k$. These points represent the wavefront distribution that bounced back from an MB and would be merged to a diffraction-limited distribution during DAS beamforming. For GT frame rendering, we isolate the tip of each wavefront in the transducer channel data by,
\begin{align}
    y_i^\star = \underset{k}{\operatorname{arg\,max}} \left\{y_{i,k}'\right\} \,,
    \quad \text{and} \quad 
    z_i^\star = \underset{k}{\operatorname{min}}\left\{z_{i,k}'\right\} \,,
\end{align}
which serve as training labels $\mathbf{p}_i^\star=\left[y_i^\star,z_i^\star,1\right]^\intercal$ for the channel data. 

\textbf{Inverse Point Transformation:}
After localization, we wish to remap channel coordinates back to \mbox{B-mode} points for comparison. An analytical inverse turns out to be infeasible due to the Euclidean distance reduction in~\eqref{eq:project}. Instead, we reverse the mapping by an affine transformation,

\begin{align}
    \begin{bmatrix}
    y_i \\
    z_i \\
    \end{bmatrix}
    &=
    \begin{bmatrix}
    a_{11} & a_{12} & a_{13} \\
    a_{21} & a_{22} & a_{23} \\
    \end{bmatrix}
    \begin{bmatrix}
    y_i^\star \\
    z_i^\star \\
    1 \\
    \end{bmatrix}
    \, ,
\end{align}
where $\left(a_{11}, a_{12}, a_{13}, a_{21}, a_{22}, a_{23}\right)$ make up the affine matrix $\mathbf{A}\in\mathbb{R}^{2\times3}$. The coefficients $(a_{11}, a_{12}, a_{21}, a_{22})$ take care of the scaling and shearing while $(a_{13}, a_{23})$ translate points. We employ the \mbox{Levenberg-Marquardt} scheme for an iterative \mbox{least-squares} optimization of $\mathbf{A}$ using, 
\begin{align} \underset{\mathbf{A}}{\min} \, \,
\left\{\rVert\mathbf{A}\mathbf{p}_i^\star-\mathbf{p}_i\rVert_2^2\right\} \, ,
\end{align}
as the objective function. For the regression, we rely on synthetic random data points $\mathbf{p}_i$ in \mbox{B-mode} space with indices $i\in\{1,2,\dots,N\}$ while $N\gg6$. The synthetic coordinates are projected to transducer channel points by \eqref{eq:project} such that $\mathbf{A}$ can be acquired once in advance and independent of training and inference. Note that coherent compounding requires to estimate $\mathbf{A}$ for each direction of wave transmission. \par
We fuse points over compounded waves via density-based clustering (DBSCAN)~\cite{ester1996density}. 
The distance criterion is chosen to be slightly greater than an expected localization error with a maximum of 0.6 wavelength units. The minimum cluster size is 1 for MBs only appearing at a single wave transmission.
\section{Materials}
\label{sec:materials}

\subsection{Implementation}
The herein used DNNs are implemented and trained in PyTorch using a single Nvidia RTX 3090. Inference is performed on an Nvidia RTX 2080 with batch size 1 to measure computation times for each incoming frame. Given a $256\times128$ input resolution with scale $R=12$ for inference, SG-SPCN occupies less than 1.5 GigaByte of GPU memory and thus suits embedded devices. Our code including SOTA methods is made available as an online respository\footnote{Access to our code repository at \href{https://github.com/hahnec/rf-ulm}{https://github.com/hahnec/rf-ulm}\label{foot:code}}.

Spatio-temporal filtering is a key pre-processing step applied prior to network inference to remove reflections from static scatterers such as tissue surfaces or bone structures. In this study, we incorporate Singular Value Decomposition (SVD) in conjunction with a temporal bandpass filter as used in~\cite{heiles2022pala} for an effective and fast removal of reflectors other than MBs. 
However, one of the key implications of our proposed method is that temporal filtering has to be applied on channel data whereas common ULM pipelines conduct temporal filtering after beamforming~\cite{demene2015spatiotemporal,baranger2018adaptive,heiles2022pala}.

For benchmark analysis on \textit{in silico} data, our network was trained with $R=8$ to ensure fair comparison with the original U-Net-based implementation~\cite{van2020super}. However, the PALA study~\cite{heiles2022pala} presents results at $R=10$, which we wish to incorporate for comparison. This discrepancy means that NMS-based networks trained at $R=8$ may produce outputs where certain pixel coordinates are never occupied. To address this issue, we introduce additive sampling noise after quantitative analysis to mitigate coordinate quantization gaps for $R<10$, allowing for qualitative comparison with other ULM rendering methods. This coordinate noise is selected to be within half the pixel size of $R$ to preserve localization accuracy in rendered images. 
\subsection{Baseline Methods}
We compare our approach with state-of-the-art methods that utilize beamforming for MB localization using classical image approaches~\cite{song:2018}, Radial Symmetry~(RS)~\cite{loy2003fast,heiles2022pala} and deep-learning-based architectures~\cite{van2020super,liu2020deep}.
\subsubsection{Classical Localization}
We obtain the results for classical image processing techniques by employing the source code released by the authors of the PALA dataset~\cite{heiles2022pala}. 

\subsubsection{Deep Learning ULM}
Existing ULM render engines based on deep learning borrowed established architectures from the imaging domain. Sloun~\textit{et al.}~\cite{van2020super} employed a U-Net~\cite{ronneberger2015u} with modifications that are explained hereafter. Similarly, Liu~\textit{et al.}~\cite{liu2020deep} propose an mSPCN that incorporates the ESPCN from Shi~\textit{et al.}~\cite{shi2016real} as a way to leverage high resolution localization. We employ the source code of mSPCN made available on the IEEE DataPort (DOI: \href{https://dx.doi.org/10.21227/jdgd-0379}{10.21227/jdgd-0379}).
As there is no publicly available implementation of~\cite{van2020super} to date, we model and train the U-Net according to the paper description, including layer architecture with the incorporation of dropout and loss design given in~\eqref{eq:loss}. Here, $\mathbf{X}$ and $\mathbf{Y}$ now represent image and labels for \mbox{B-mode} frames, respectively.
As per~\cite{van2020super}, we set $\lambda_0=1$ and $\lambda_1=1\mathrm{e}{-2}$ for the regularized U-Net loss. The mSPCN is regularized with $\lambda_0=50$ and $\lambda_1=1$.
Employment of the U-Net model for ULM requires to upscale \mbox{B-mode} frames by factor $R$ prior to inference~\cite{van2020super}. It is important to note that the order of the \mbox{2-D} interpolation method has a significant impact on the localization accuracy. We choose the \mbox{bi-cubic} approach in this study to achieve best U-Net results. Due to the large image input size and the network's memory requirements, the U-Net training is limited to \mbox{$R=8$} in this study, which is in accordance with~\cite{van2020super}. 
Benchmarking these DNNs requires to determine single point coordinates for each MB position. Previous DNN-based studies on ULM accumulate network outputs, where each MB prediction is distributed over several pixels. For fair comparison, we apply NMS thresholding to the baseline methods to obtain MB coordinates. 

\begin{figure*}[b]
  \centering
  \begin{minipage}[b]{0.120\textwidth}
    \centering
    \includegraphics[width=\textwidth,trim=240 538 1030 238, clip]{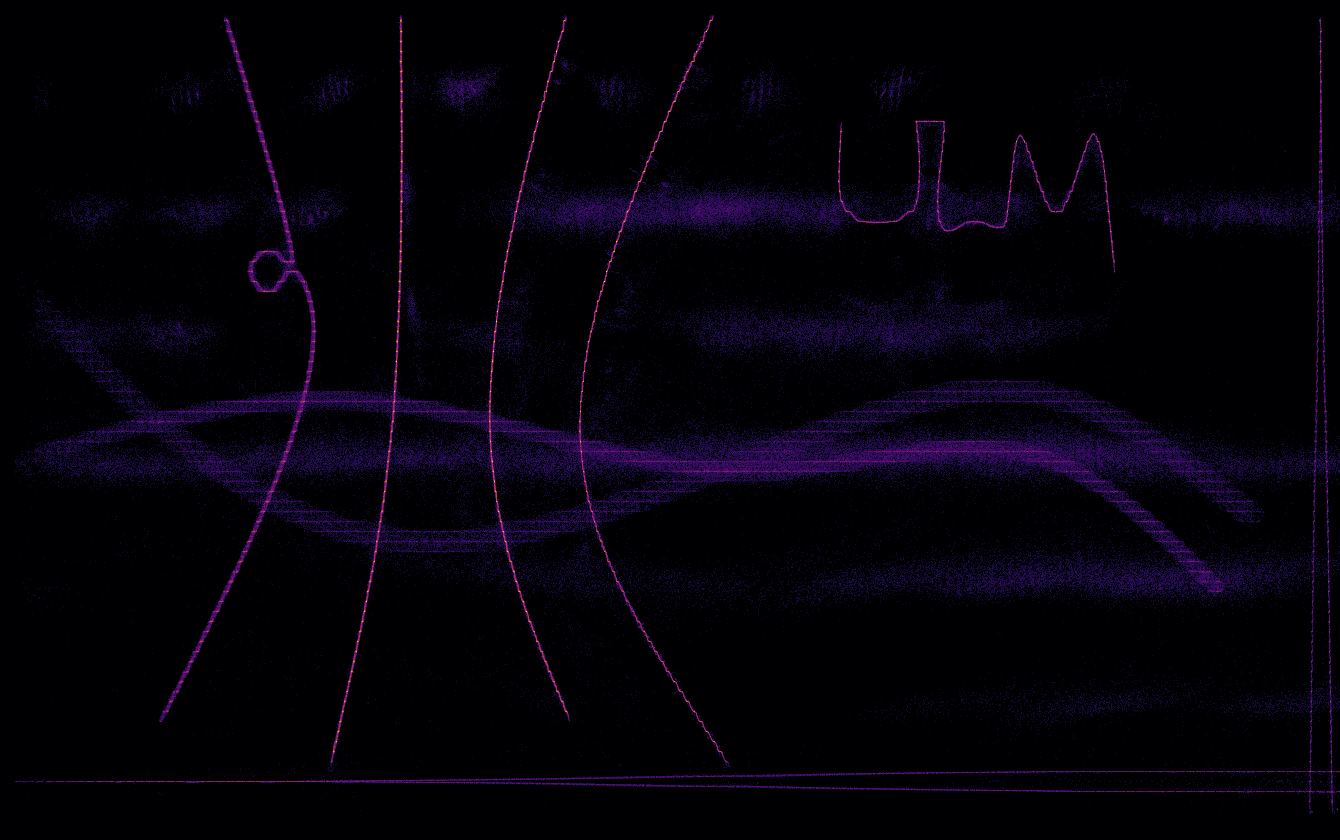}\par\vspace{1mm}
    \includegraphics[width=\textwidth,trim=310 45 980 750, clip]{./figures/pala_sim_results/pala_ulm_img_15k_gam0.9_128chs.png}\par\vspace{1mm}
    \includegraphics[width=\textwidth,trim=1240 20 0 715, clip]{./figures/pala_sim_results/pala_ulm_img_15k_gam0.9_128chs.png}\par
    \subcaption{RS~\cite{heiles2022pala}\label{fig:method2}}
  \end{minipage}
  \hfill
    \begin{minipage}[b]{0.120\textwidth}
        \centering
        \includegraphics[width=\textwidth,trim=240 538 1030 238, clip]{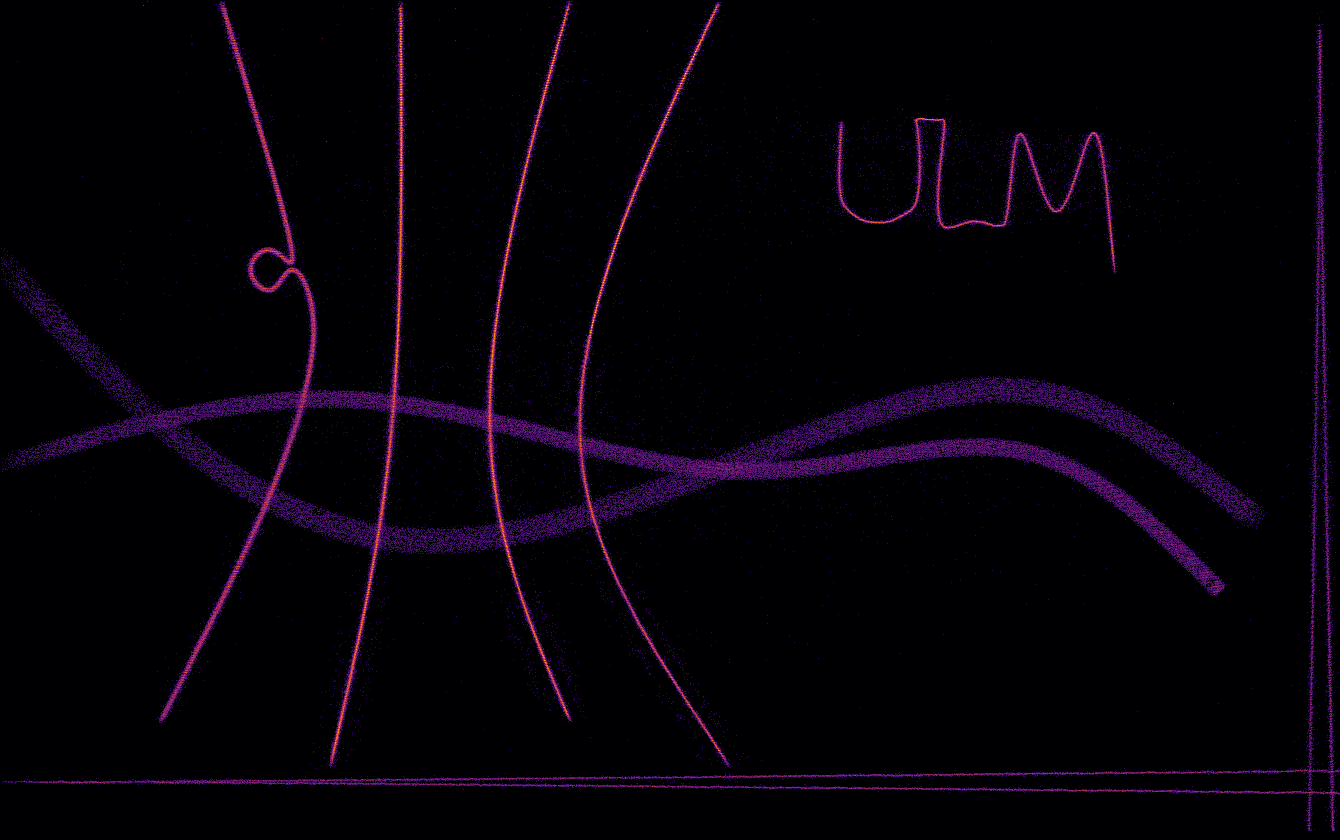}\par\vspace{1mm}
        \includegraphics[width=\textwidth,trim=310 45 980 750, clip]{./figures/pala_sim_results/gulm_ulm_img_15k_gam0.9_16chs.png}\par\vspace{1mm}
        \includegraphics[width=\textwidth,trim=1240 20 0 715, clip]{./figures/pala_sim_results/gulm_ulm_img_15k_gam0.9_16chs.png}\par
        \subcaption{G-ULM~\cite{gulm:2023}\label{fig:method3}}
    \end{minipage}
  \hfill
  \begin{minipage}[b]{0.120\textwidth}
    \centering
    \includegraphics[width=\textwidth,trim=240 538 1030 238, clip]{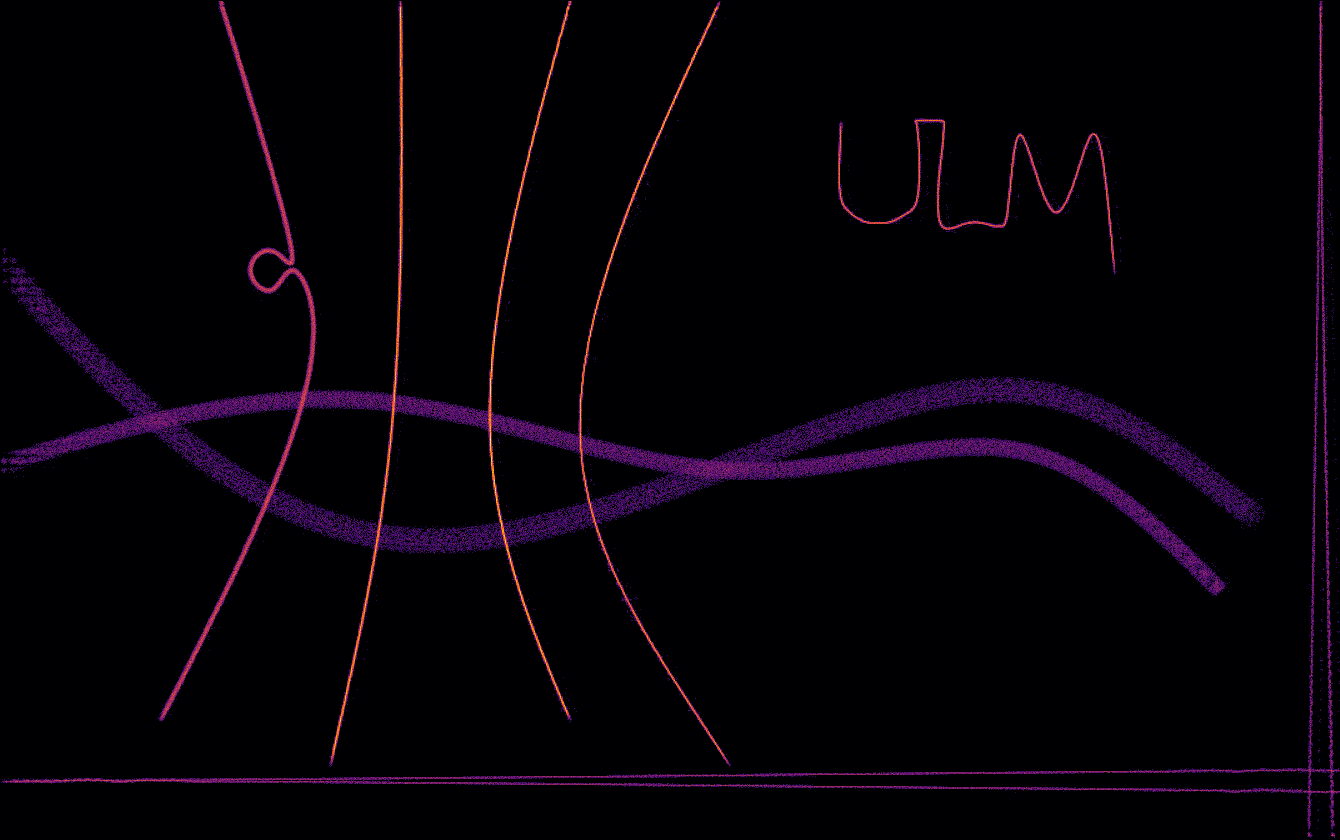}\par\vspace{1mm}
    \includegraphics[width=\textwidth,trim=310 45 980 750, clip]{./figures/pala_sim_results/unet_8x_iq.png}\par\vspace{1mm}
    \includegraphics[width=\textwidth,trim=1240 20 0 715, clip]{./figures/pala_sim_results/unet_8x_iq.png}
    \subcaption{U-Net~\cite{van2020super}\label{fig:method4}}
  \end{minipage}
  \hfill
  \begin{minipage}[b]{0.120\textwidth}
    \centering
    \includegraphics[width=\textwidth,trim=240 538 1030 238, clip]{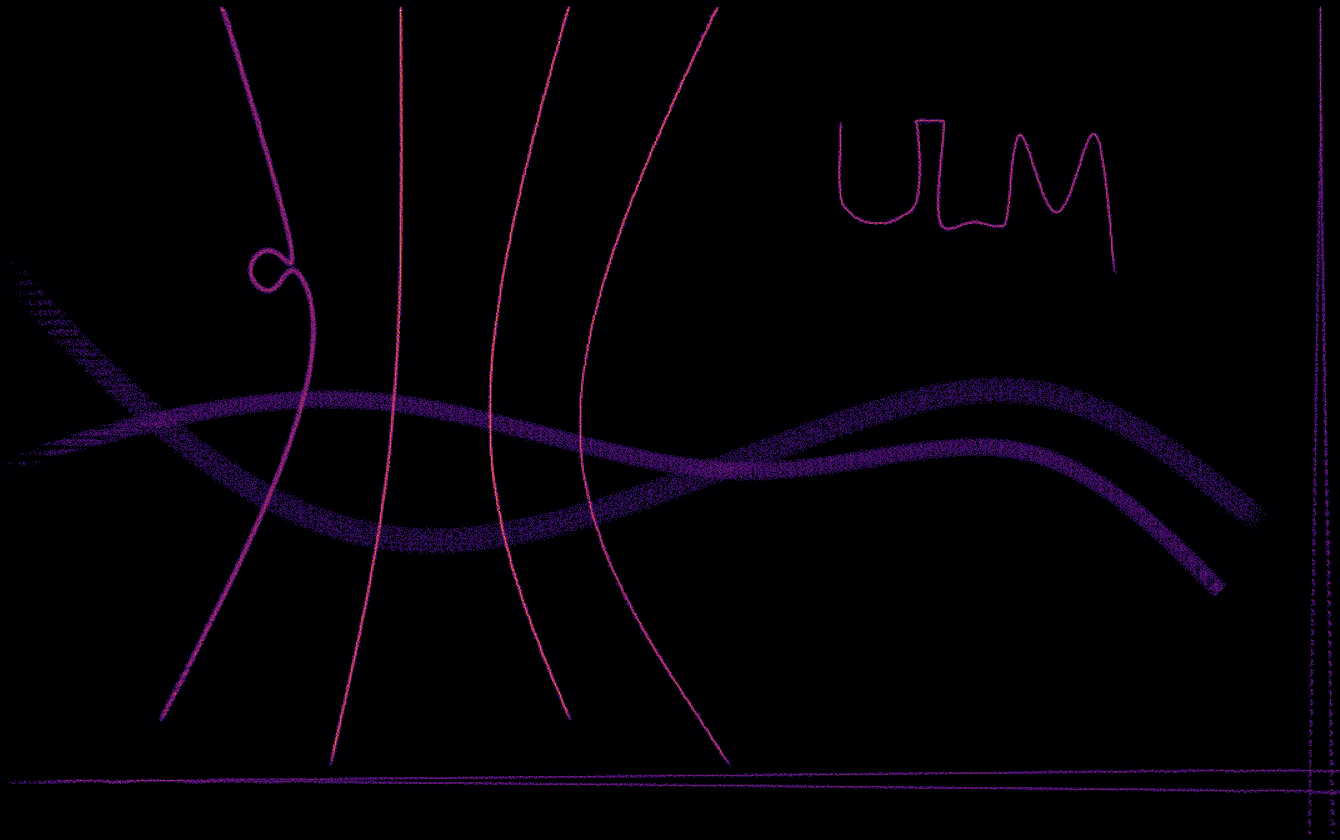}\par\vspace{1mm}
    \includegraphics[width=\textwidth,trim=310 45 980 750, clip]{./figures/pala_sim_results/mspcn_8x_iq.png}\par\vspace{1mm}
    \includegraphics[width=\textwidth,trim=1240 20 0 715, clip]{./figures/pala_sim_results/mspcn_8x_iq.png}
    \subcaption{mSPCN~\cite{liu2020deep}\label{fig:method5}}
  \end{minipage}
  \hfill
  \begin{minipage}[b]{0.120\textwidth}
    \centering
    \includegraphics[width=\textwidth,trim=240 538 1030 238, clip]{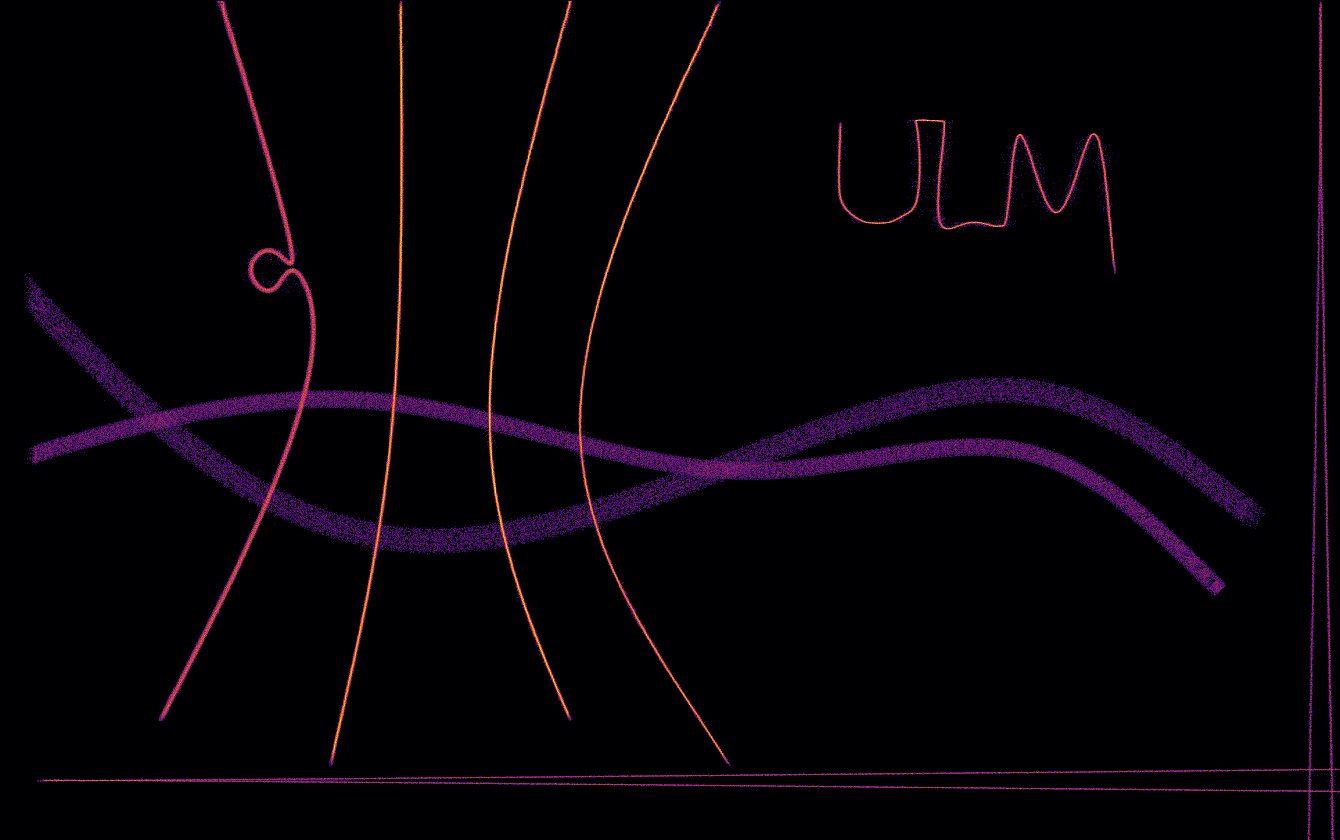}\par\vspace{1mm}
    \includegraphics[width=\textwidth,trim=310 45 980 750, clip]{./figures/pala_sim_results/sgspcn_8x_rf.png}\par\vspace{1mm}
    \includegraphics[width=\textwidth,trim=1240 20 0 715, clip]{./figures/pala_sim_results/sgspcn_8x_rf.png}
    \subcaption{
    Ours\label{fig:method6}}
  \end{minipage}
  \hfill
  \begin{minipage}[b]{0.120\textwidth}
    \centering
    \begin{tikzpicture}
        \node[inner sep=0pt] at (0,0) {
        \includegraphics[width=\textwidth]{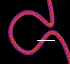}};
        \node[white] at (.525,-.575) {$1.7\lambda$};
    \end{tikzpicture}
    \par\vspace{1mm}
    \begin{tikzpicture}
        \node[inner sep=0pt] at (0,0) {
        \includegraphics[width=\textwidth]{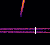}};
        \node[white] at (.5,-.7) {$0.7\lambda$};
    \end{tikzpicture}
    \par\vspace{1mm}
    \begin{tikzpicture}
        \node[inner sep=0pt] at (0,0) {
        \includegraphics[width=\textwidth]{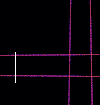}};
        \node[white] at (-.575,-.9) {$2.9\lambda$};
    \end{tikzpicture}
    \subcaption{GT\label{fig:method1}}
  \end{minipage}
  \begin{minipage}[b]{0.233\textwidth}
    \raggedright
    {\footnotesize Counts [\#]}\hfill
    \vspace{-1mm}
    \includegraphics[width=\textwidth]{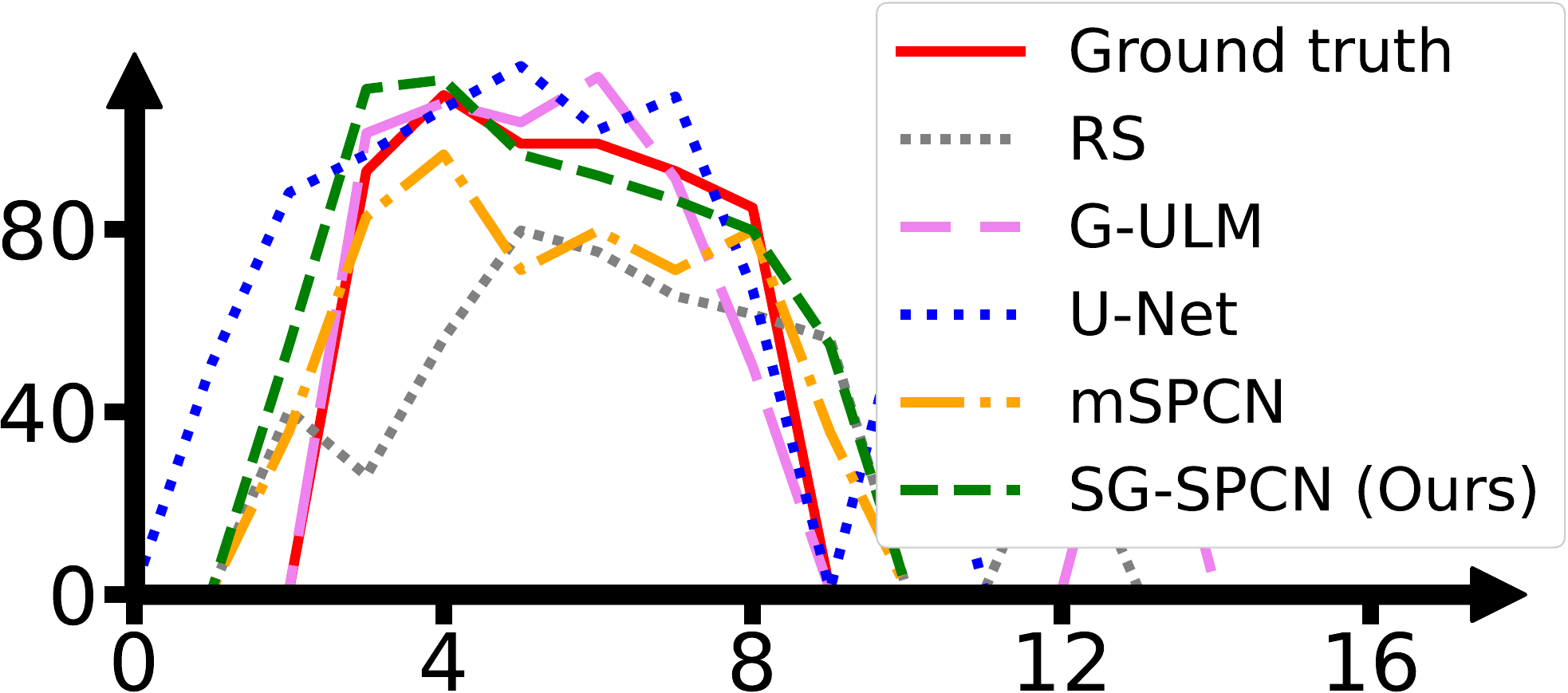}\par\vspace{3mm}
    \includegraphics[width=\textwidth]{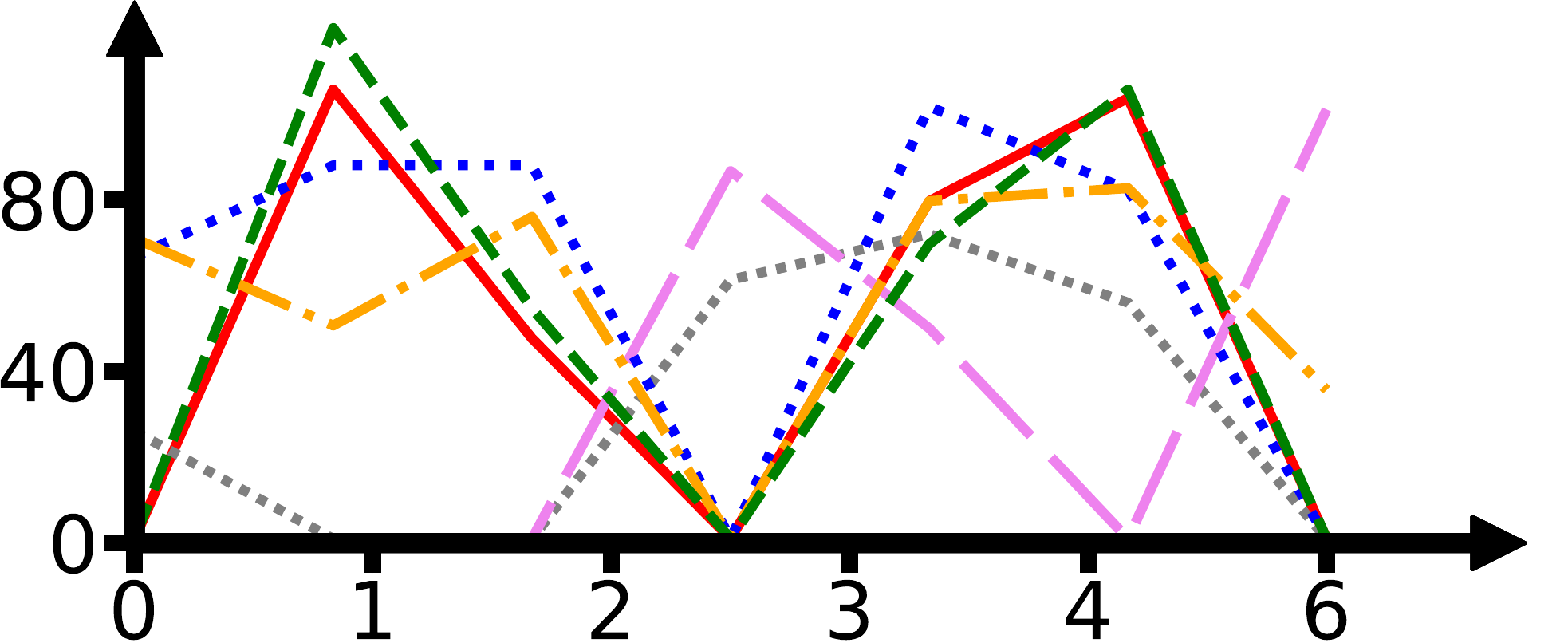}\par\vspace{3.5mm}
    \includegraphics[width=\textwidth]{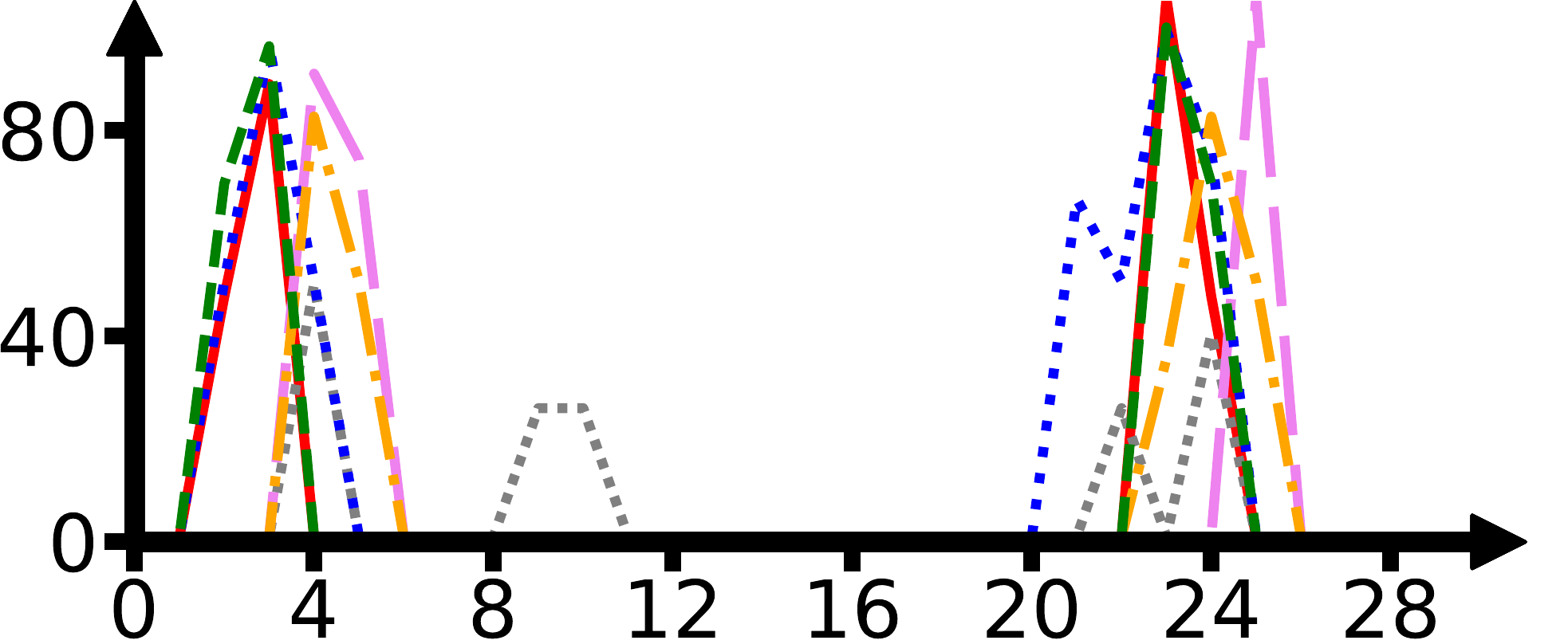}\par\vspace{0mm}
    \par\vspace{-1mm}
    \centering
    {\footnotesize Spatial dimension [px]}
    \par\vspace{-.5mm}
    \subcaption{Cross-sections\label{fig:cross-section}}
  \end{minipage}
    \caption{\textbf{In silico ULM regions} from  Table~\ref{tab:benchmark} for localization assessment. The methods in (\subref{fig:method2})-(\subref{fig:method3}) are deterministic approaches whereas (\subref{fig:method4})-(\subref{fig:method6}) are based on deep learning models with scale $R=8$. All images are rendered with $R=10$ to emphasize deviations from the GT. The results in (\subref{fig:method3}) and (\subref{fig:method6}) are generated in absence of computational beamforming. Our SG-SPCN network renderings are found in (\subref{fig:method6}). Cross-sections are highlighted as white bars in (\subref{fig:method1}) and depicted in (\subref{fig:cross-section}).} 
    \label{fig:render_ulm_crop}
\end{figure*}

\subsection{Datasets}
%
\subsubsection{In Silico Angiography Dataset}
We employ the available \textit{in silico} data from the PALA study~\cite{heiles2022pala} for training and benchmark testing. The Verasonics Research Ultrasound Simulator generated the channel data and B-mode frames within a \mbox{7 × 14.9 mm$^2$} area, utilizing 3 tilted plane waves (-5, 0, and 5 degrees) from a 15.6~MHz linear probe with 128 elements at 0.1~mm pitch. Realistic MB motion is simulated using 11 manually shaped tubes following the Poiseuille's law velocity profile. The speed of sound amounts to 1540 m/s. To create a ground truth map~$\mathbf{Y}$, we set the sample positions at scatterer coordinates (i.e.,~$\mathbf{p}^{\star}_i$) to one. 
It is important to note that the PALA dataset features \mbox{B-mode} frames with $143\times84$ pixels originating from the $128\times256$ I/Q channels~\cite{heiles2022pala}. This beamforming transformation involves upsampling in the lateral domain and efficient sample removal in the depth dimension. Based on these frame sizes, the inputs are randomly cropped to square patches of 128 I/Q channel data samples and 64 B-mode pixels.
As part of supervised learning, we partition the PALA \textit{in silico} dataset into testing sequences 1-15 synthetic image rendering and training/validation sequences 16-20, using a 0.9 split ratio, resulting in 4500 training frames. This dataset can be found at \href{https://doi.org/10.5281/zenodo.4343435}{doi.org/10.5281/zenodo.4343435}.

\subsubsection{In Vivo Dataset}
In the \textit{in vivo} rat brain perfusion datasets, Sprague-Dawley rats (8–10 weeks old) were used, adhering to ethical guidelines. The rats were acclimated for at least a week before surgery, provided with water, and a commercial pelleted diet. After anesthesia induction with isoflurane, craniotomy surgery was performed to create a 14~mm-wide window for imaging. Sonovue MBs were injected continuously or as bolus. The \textit{in vivo} data was captured with 128 elements at 0.1~mm pitch, 15.6~MHz central frequency (67\% relative bandwidth), 1000~Hz frame rate, and 5 tilted plane waves (–6, –3, 0, 3, and 6 degrees). 
The baseline methods utilize \textit{in vivo} B-mode frames from DAS with axial sampling equivalent to channel data, which were recorded in the PALA study~\cite{heiles2022pala} and made available at \href{https://doi.org/10.5281/zenodo.7883227}{doi.org/10.5281/zenodo.7883227}.

\subsection{Metrics}

We assess our results using established field metrics~\cite{liu2020deep,heiles2022pala,hahne:23:ius}. To measure localization accuracy, we calculate the minimum Root Mean Squared Error (RMSE) between estimated and GT positions. Following the method by~\cite{heiles2022pala}, we consider RMSEs smaller than a quarter of the wavelength as true positives, contributing to the overall RMSE across frames. Larger RMSEs lead to classifying the estimated position as a false positive. GT locations without an estimate within the wavelength threshold are marked as false negatives. We evaluate detection reliability using the Jaccard Index, which considers true positives, false positives and false negatives, offering a robust performance measure for each algorithm. For image quality analysis, we utilize the Structure Similarity Index Measure (SSIM)~\cite{wang2004image}. We further report weight parameter count and inference time for each model with batch size 1.
\begin{table*}[!h]
	\centering
	\caption{Localization results from 15000 test frames of the PALA dataset~\cite{heiles2022pala} where each network is trained with \mbox{$R=8$} for fair comparison. Metrics are reported as mean±std. where applicable with units provided in brackets. Vertical arrows indicate direction of better scoring and $\text{T}_{\text{DAS}}$ denotes the DAS 
    beamforming interval and $\text{T}_R$ the upsampling for each frame.\label{tab:benchmark}}
    \begin{tabularx}{\textwidth}{ 
         >{\raggedright\arraybackslash}p{9.5em} 
         >{\raggedright\arraybackslash}p{4em} 
         >{\raggedleft\arraybackslash}p{2.75em} 
         >{\raggedleft\arraybackslash}p{8em} 
         >{\raggedleft\arraybackslash}p{6em} 
         >{\raggedleft\arraybackslash}p{6em} 
         >{\raggedleft\arraybackslash}p{6em} 
         >{\raggedleft\arraybackslash}p{10em} 
         }
\toprule
Method & Input & Waves & {RMSE [$\lambda/10$]}~$\downarrow$ & {Jaccard~[\%]}~$\uparrow$ & {SSIM~[\%]}~$\uparrow$ & Weights [\#]~$\downarrow$ & Frame Time [ms]~$\downarrow$ \\
\midrule
Lanczos~\cite{heiles2022pala} & B-mode & 3 & $1.524\pm0.175$ & $38.688$ & $75.870$ & $0$ & $\text{T}_{\text{DAS}}$ + $0.382$ $\times$ 1$\mathrm{e}$3 \\
2-D Gauss Fit~\cite{song:2018} & B-mode & 3 & $1.240 \pm 0.162$ & $51.342$ & $73.930$ & $0$ & $\text{T}_{\text{DAS}}$ + $3.782$ $\times$ 1$\mathrm{e}$3 \\
RS~\cite{heiles2022pala} & B-mode & 3 & $1.179 \pm 0.172$ & $50.330$ & $72.170$ & $0$ & $\text{T}_{\text{DAS}}$ + $0.099$ $\times$ 1$\mathrm{e}$3 \\ 
G-ULM~\cite{gulm:2023} & RF$\to$I/Q & 1 & $0.967\pm0.109$ & $78.618$ & $92.020$ & $0$ & $3.747$ $\times$ 1$\mathrm{e}$3 \\
\midrule
U-Net~\cite{van2020super} + NMS & B-mode & 3 & $0.580 \pm 0.081$ & $90.192$ & $93.700$ & $12982849$ & $\text{T}_{\text{DAS}}$ + $\text{T}_{R}$ + $54.454$ \\
{mSPCN~\cite{liu2020deep}} + NMS & B-mode & 3 & $0.696 \pm 0.097$ & $85.406$ & $92.829$ & $453568$ & $\text{T}_{\text{DAS}}$ + $2.715$ \\
{mSPCN~\cite{liu2020deep}} + NMS & RF$\to$I/Q & 3 & $1.095 \pm 0.192$ & $57.056$ & $89.361$ & $453568$ & $18.280$ \\
{SG-SPCN [Ours]} & B-mode & 3 & $0.627 \pm 0.092$ & $89.519$ & $93.783$ & $658496$ & $\text{T}_{\text{DAS}}$ + $3.258$ \\
{SG-SPCN [Ours]} & RF$\to$I/Q & 1 & $0.564
 \pm 0.091$ & $85.894$ & $94.012$ & $658496$ & $6.728$ \\
{SG-SPCN [Ours]} & RF$\to$I/Q & 3 & $0.412 \pm 0.084$ & $88.106$ & $94.316$ & $658496$ & $16.752$ \\
\midrule
U-Net~\cite{van2020super} + NMS + RS & B-mode & 3 & $0.415 \pm 0.088$ & $90.320$ & $93.261$ & $12982849$ & $\text{T}_{\text{DAS}}$ + $\text{T}_{R}$ + $83.593$ \\
{SG-SPCN [Ours] + RS} & RF$\to$I/Q & 3 & $0.322 \pm 0.086$ & $88.190$ & $94.160$ & $658496$ & $33.565$ \\
\bottomrule
\end{tabularx}
\end{table*}
\section{Results}
\label{sec:results}
\subsection{In Silico Benchmark}
%
This study presents a manifold evaluation of our trained network's performance for ULM rendering. At first, we conduct benchmark comparisons using available GT data~\cite{heiles2022pala}, with both qualitative and quantitative assessments presented in Fig.~\ref{fig:render_ulm_crop} and Table~\ref{tab:benchmark}, respectively. Although we use \mbox{$R=8$}, note that the images in Fig.~\ref{fig:render_ulm_crop} are rendered at scale 10 to guarantee the highest precision for the GT frame. One striking observation in Table~\ref{tab:benchmark} is the accuracy achieved by our RF-ULM network, \mbox{SG-SPCN}, as evidenced by a mean RMSE improvement of more than 20\% compared to~\cite{van2020super} as the second-best approach. To enable a direct comparison with \mbox{B-mode} counterparts, we train network architectures using both \mbox{B-mode} and channel input data. 
Notably, our RF-ULM network outperforms \mbox{B-mode-based} networks, and several factors contribute to this:
First, RF channel data contains wavefront distributions that provide richer spatial information, enabling the network to make more accurate predictions based on geometric shapes. In particular, the hyperbolic curvatures present in RF channels assist the network in precisely locating the tips of arriving wavefronts. Furthermore, our analysis reveals variability in localization accuracy for \mbox{RF-based} networks, with outstanding results in regions closer to the transducer probe such as the bottom row of Fig.~\ref{fig:render_ulm_crop}. %

On the contrary, the U-Net approach~\cite{van2020super} exhibits a slightly higher Jaccard index, primarily attributed to its roughly 20 times more weights. Besides, the unique upscaling of input frames prior to inference is an additional process not adopted by other methods as it imposes high complexity demands resulting in a more than 4 times slower computation.
%

Considering the need for temporal data in ULM, we recognize the stringent time constraints involved. 
To illustrate, a theoretical minimum of 1000 input frames and a hypothetical localization interval of 1~ms per frame will ideally take a few seconds to render the full image given that ULM requires additional pre-processing such as temporal filters. 
As seen in Table~\ref{tab:benchmark}, combining the mSPCN model with \mbox{B-mode} images enables rapid computation, requiring less than a minute for 15,000 frames. However, this requires fast DAS beamforming with $\text{T}_{\text{DAS}}\lesssim 1$~ms. Our beamforming implementation with PyTorch takes about 100~ms per frame on the Nvidia RTX 2080~\cite{hahne2024learning}. Although there is a potential for speed optimization, beamforming poses a significant time bottleneck, which is replaced in RF-ULM by our geometric transform taking about 0.8~ms with the NumPy library. 
This enables our RF-based network to achieve superior ULM image scores from direct localization with a competitive computation time of about 100 seconds for a single wave emission and 4 minutes for 3 compounded waves. Feeding multiple emissions into a network and then merging points through subsequent clustering comes with a computational overhead.

\begin{figure}[h!]
    \centering
    \includegraphics[width=\linewidth]{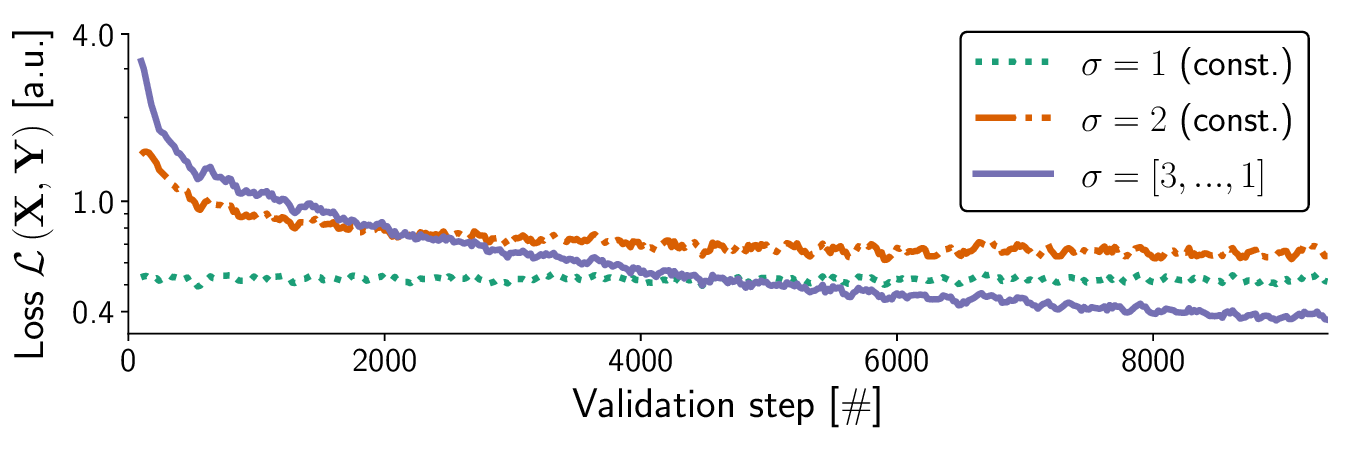}
    \caption{\textbf{Validation loss} of varying $\sigma$ for \mbox{SG-SPCN} at $R=12$. For clear visibility, curves undergo a 90-window rolling mean.\label{fig:loss}}
\end{figure}

When analyzing our proposed $\sigma$-decrement, we observe a steeper loss progression during training as shown in Fig.~\ref{fig:loss}. Our approach proves advantageous, yielding a lower final validation loss compared to training with a constant $\sigma$.

\subsection{Ablation Study}
\label{sec:ablation}

We examine how our proposed \mbox{semi-global} module influences the performance of RF-based ULM. Our \mbox{SG-SPCN} model is built upon the mSPCN~\cite{liu2020deep} and nearly identical except for the inclusion of the \mbox{semi-global} block. 
%
%
The effectiveness of our \mbox{semi-global} scaling module becomes apparent as one examines the notable improvements when contrasting the rows of mSPCN and \mbox{SG-SPCN} in Table~\ref{tab:benchmark} for RF-data inputs.
%
Similarly, our architecture achieves a significant improvement for \mbox{B-mode} inputs, which is seen by comparing the \mbox{B-mode} rows of the mSPCN and \mbox{SG-SPCN}, respectively. 
Despite the slight increase in processing time and parameters weights, this experimental quantification underscores our framework's advantages and potential in real data scenarios. 
\subsection{In Vivo Comparison}
For a realistic examination, we present an experimental analysis using baseline methods and \textit{in vivo} data that contains the vascular structure of two different rat brains\footnote{\href{https://doi.org/10.5281/zenodo.7883227}{doi.org/10.5281/zenodo.7883227}\label{foot:invivo}}. 
Due to the memory requirements of the U-Net approach, we provide U-Net results at $R=8$ as proposed by Sloun~\textit{et al.}~\cite{van2020super}. 
To show the full capacity of our pipeline, we train SG-SPCN with $R=12$ using B-mode and RF frames, respectively. 

Figure~\ref{fig:rat18} depicts \textit{in vivo} results including the deterministic RS~\cite{heiles2022pala} for comparison. Closer inspection reveals that our proposed RF-framework renders high quality ULM images for real-world scenarios. 
The reason for this lies in our pipeline's ability to robustly detect true positive wavefronts with high geometric precision, which is reflected by the outstanding contrast in Fig.~\ref{fig:rat18:sgspcnrf}. 
Another component for this achievement is the introduced NMS, whose effectiveness is demonstrated by the improved sharpness and contrast.
Conversely, choosing $R>8$ for the dataset's B-mode images begins to present challenges for other DNN pipelines. We attribute this limitation to the constrained spatial extent and varying shape of MBs in B-mode images, which makes it difficult for networks to learn and predict locations at such a fine level of detail (see Fig.~\ref{fig:sub:bmode_hires}). %
Using a U-Net becomes impractical due to the considerable complexity demands imposed by the $1716$ by $2016$ scaled image size of 96,000 frames. 
%

\newcommand{\overlayRectangle}[4]{
    \vspace{-.4cm}
    \begin{tikzpicture}[remember picture,overlay]
    \begin{scope}
        \fill [fill=mblue, fill opacity=.5] (#1,#2) rectangle ++ (#3*#2/#4,-0.5);
        \draw [line width=0.25em,color=white] (#1,#2) -- ++ (#3*#2/#4,0);
        \node[below,inner sep=0.225em,font=\footnotesize,color=white] at (#1+13,#2-2.5) {\SI{1}{\milli\meter}};
    \end{scope}
    \end{tikzpicture}
}

\begin{figure*}
    \centering
    \begin{minipage}[b]{.49\linewidth}
        \resizebox{\textwidth}{!}{
            \begin{tikzpicture}[x=6cm, y=6cm, spy using outlines={every spy on node/.append style={smallwindow}}]
            \node[anchor=south,scale=1.5] (FigA) at (0,0) {
            \includegraphics[height=2cm,trim=5cm 12cm 12.5cm 1.0cm,clip]{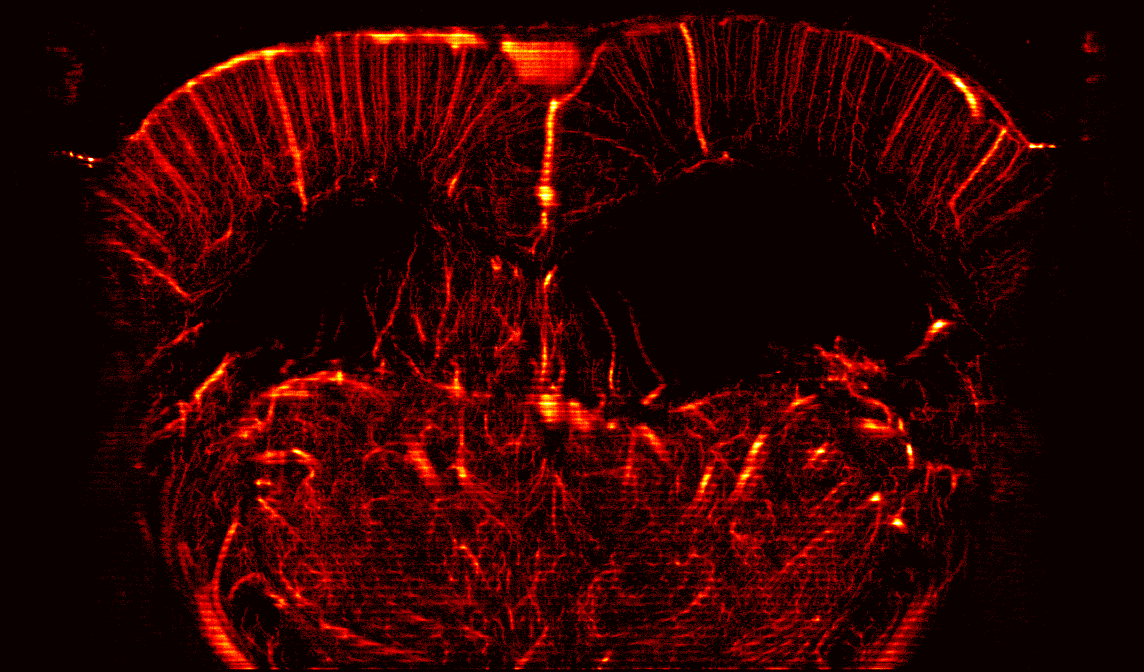}
            };
            \spy [closeup,magnification=3] on ($(FigA)+(-0.285,+0.05)$) in node[largewindow,anchor=north east] at ($(FigA.north west) - (0,0.02)$);
            \spy [closeup,magnification=2] on ($(FigA)+(+0.095,+0.022)$) in node[largewindow,anchor=east]       at ($(FigA.west)$);
            \spy [closeup,magnification=2] on ($(FigA)+(-0.435,-0.075)$) in node[largewindow,anchor=south east] at ($(FigA.south west) + (0,0.02)$);
            \end{tikzpicture}
        }
        \overlayRectangle{7.45cm}{1.25cm}{.375cm}{.5cm}
        \subcaption{U-Net~\cite{van2020super} + NMS from B-mode frames\label{fig:rat20:unet}}
        \vfill
    \end{minipage}
    \begin{minipage}[b]{.49\linewidth}
        \resizebox{\textwidth}{!}{
            \begin{tikzpicture}[x=6cm, y=6cm, spy using outlines={every spy on node/.append style={smallwindow}}]
            \node[anchor=south,scale=1.5] (FigA) at (0,0) {
    	   \includegraphics[height=2cm,trim=7.5cm 18cm 18.75cm 1.5cm,clip]{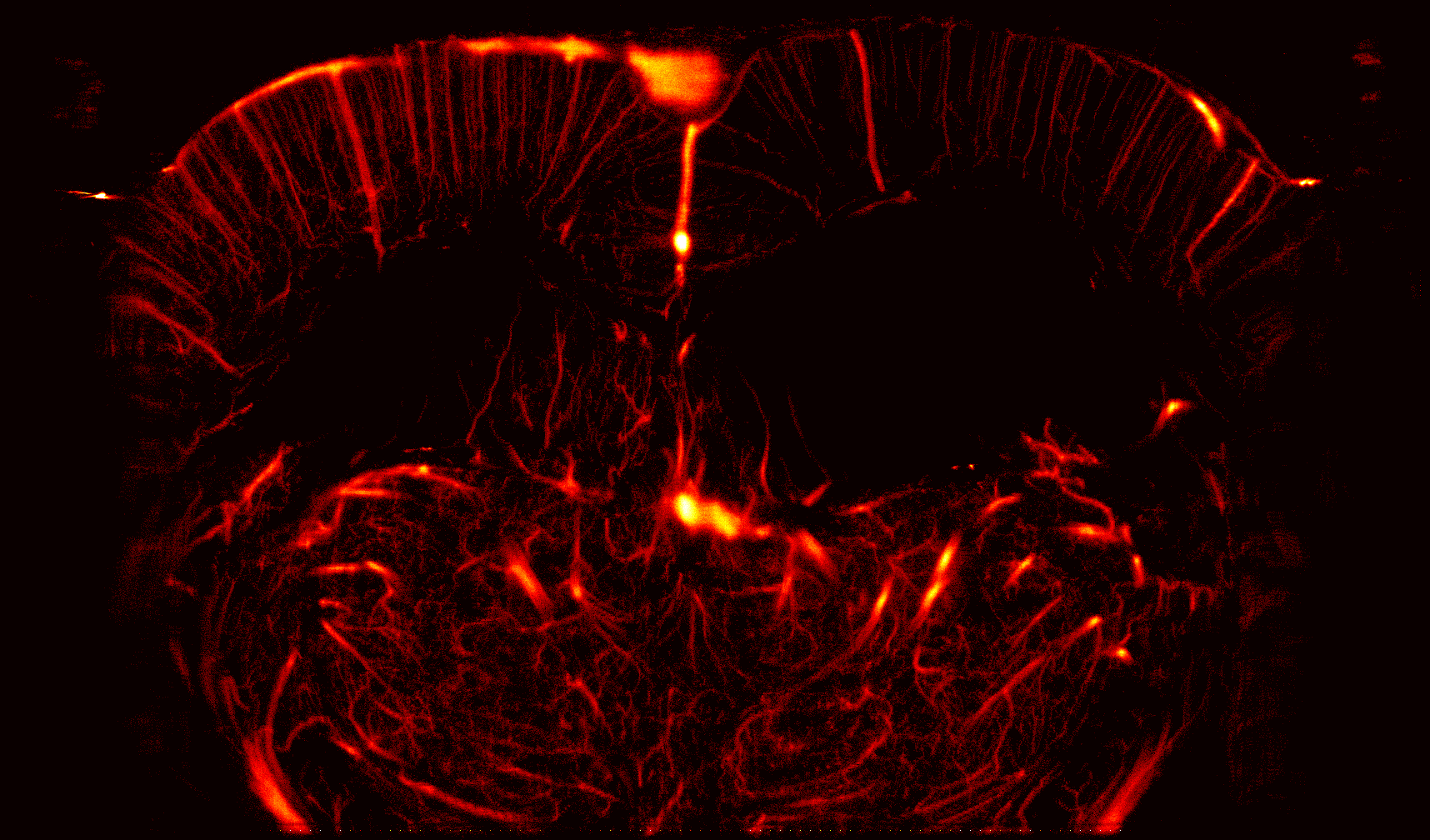}};
            \spy [closeup,magnification=3] on ($(FigA)+(-0.285,+0.05)$) in node[largewindow,anchor=north east] at ($(FigA.north west) - (0,0.02)$);
            \spy [closeup,magnification=2] on ($(FigA)+(+0.095,+0.022)$) in node[largewindow,anchor=east]       at ($(FigA.west)$);
            \spy [closeup,magnification=2] on ($(FigA)+(-0.435,-0.075)$) in node[largewindow,anchor=south east] at ($(FigA.south west) + (0,0.02)$);
            \end{tikzpicture}
        }
        \overlayRectangle{7.45cm}{1.25cm}{.375cm}{.5cm}
        \subcaption{Radial Symmetry~\cite{heiles2022pala} from B-mode frames\label{fig:rat20:rs}}
	\end{minipage}
    \begin{minipage}[b]{.49\linewidth}
        \resizebox{\textwidth}{!}{
            \begin{tikzpicture}[x=6cm, y=6cm, spy using outlines={every spy on node/.append style={smallwindow}}]
            \node[anchor=south,scale=1.5] (FigA) at (0,0) {
            \includegraphics[height=2cm,trim=7.5cm 18cm 18.75cm 1.5cm,clip]{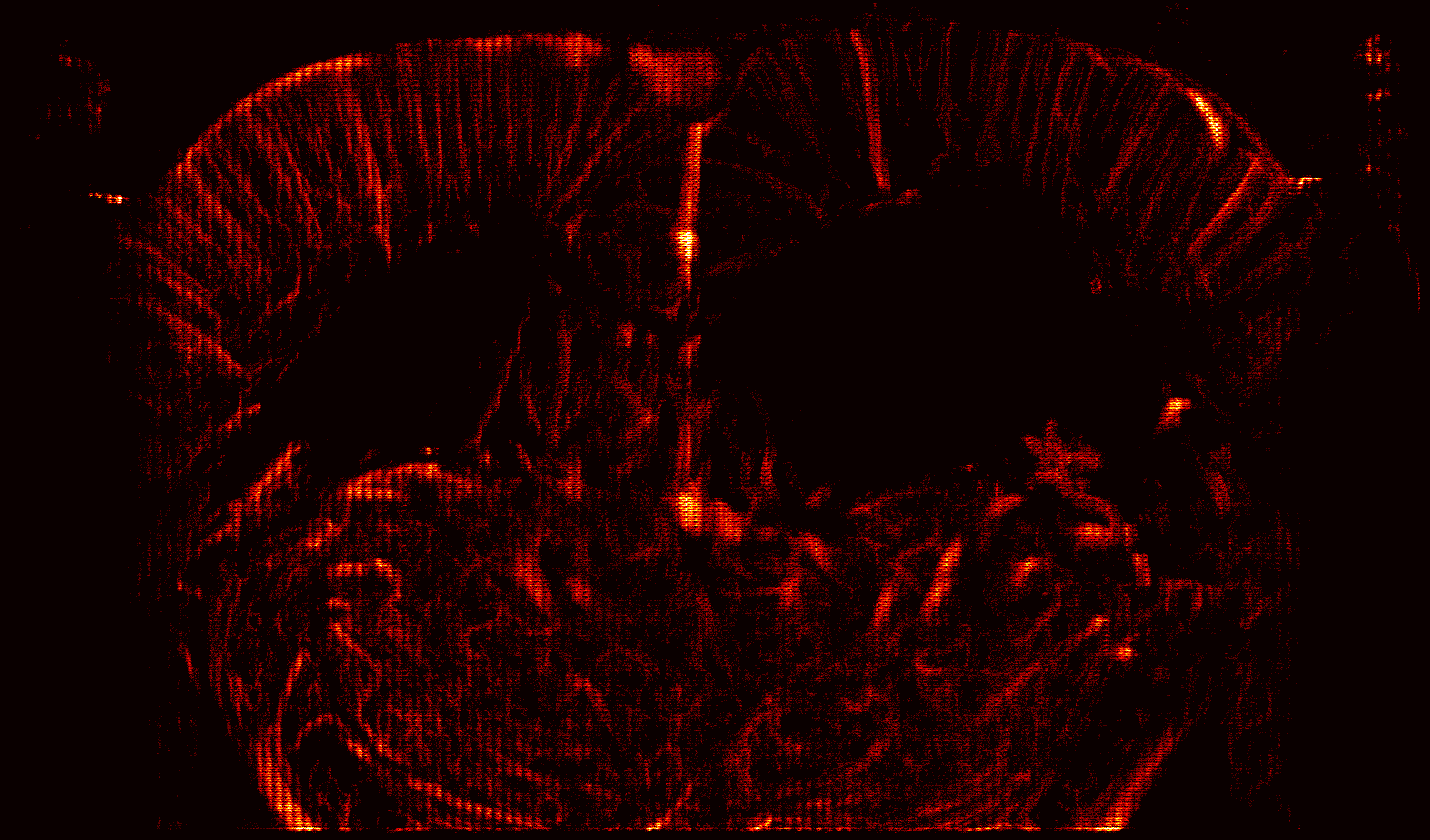}};
            \spy [closeup,magnification=3] on ($(FigA)+(-0.285,+0.05)$) in node[largewindow,anchor=north east] at ($(FigA.north west) - (0,0.02)$);
            \spy [closeup,magnification=2] on ($(FigA)+(+0.095,+0.022)$) in node[largewindow,anchor=east]       at ($(FigA.west)$);
            \spy [closeup,magnification=2] on ($(FigA)+(-0.435,-0.075)$) in node[largewindow,anchor=south east] at ($(FigA.south west) + (0,0.02)$);
            \end{tikzpicture}
        }
        \overlayRectangle{7.45cm}{1.25cm}{.375cm}{.5cm}
        \subcaption{SG-SPCN + NMS from B-mode frames\label{fig:rat20:sgspcnbmode}}
        \vfill
    \end{minipage}
    \begin{minipage}[b]{.49\linewidth}
        \resizebox{\textwidth}{!}{
            \begin{tikzpicture}[x=6cm, y=6cm, spy using outlines={every spy on node/.append style={smallwindow}}]
            \node[anchor=south,scale=1.5] (FigA) at (0,0) {
    	   \includegraphics[height=2cm,trim=7.5cm 18cm 18.75cm 1.5cm,clip]{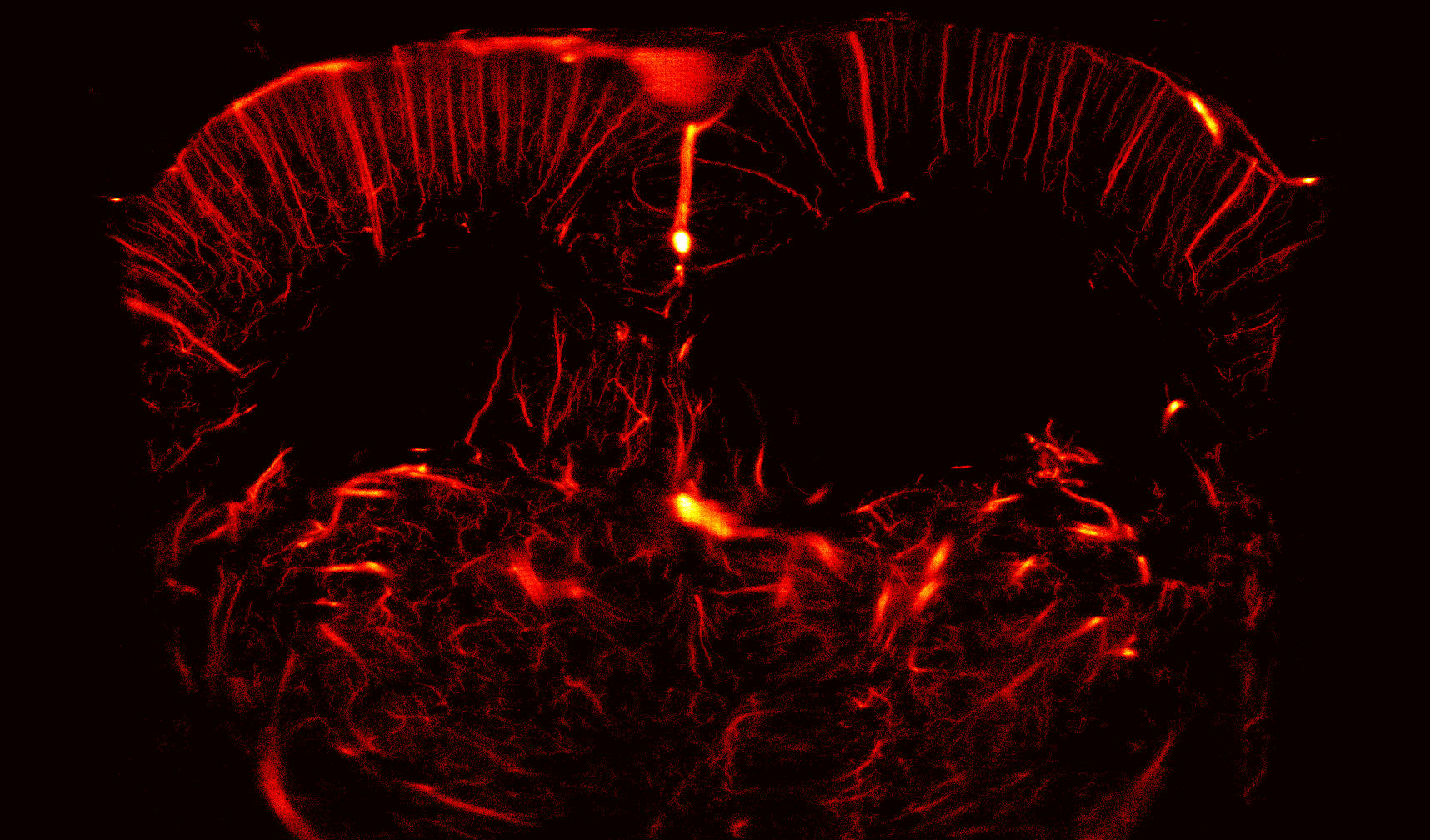}};
            \spy [closeup,magnification=3] on ($(FigA)+(-0.285,+0.05)$) in node[largewindow,anchor=north east] at ($(FigA.north west) - (0,0.02)$);
            \spy [closeup,magnification=2] on ($(FigA)+(+0.095,+0.022)$) in node[largewindow,anchor=east]       at ($(FigA.west)$);
            \spy [closeup,magnification=2] on ($(FigA)+(-0.435,-0.075)$) in node[largewindow,anchor=south east] at ($(FigA.south west) + (0,0.02)$);
            \end{tikzpicture}
        }
        \overlayRectangle{7.45cm}{1.25cm}{.375cm}{.5cm}
        \subcaption{SG-SPCN + NMS from RF$\to$I/Q frames\label{fig:rat20:sgspcnrf}}
	\end{minipage}
    \begin{minipage}[b]{.49\linewidth}
    \resizebox{\textwidth}{!}{
        \begin{tikzpicture}[x=6cm, y=6cm, spy using outlines={every spy on node/.append style={smallwindow}}]
        \node[anchor=south,scale=1.5] (FigA) at (0,0) {
        \includegraphics[height=2cm,trim=6cm 11.67cm 12cm 1.67cm,clip]{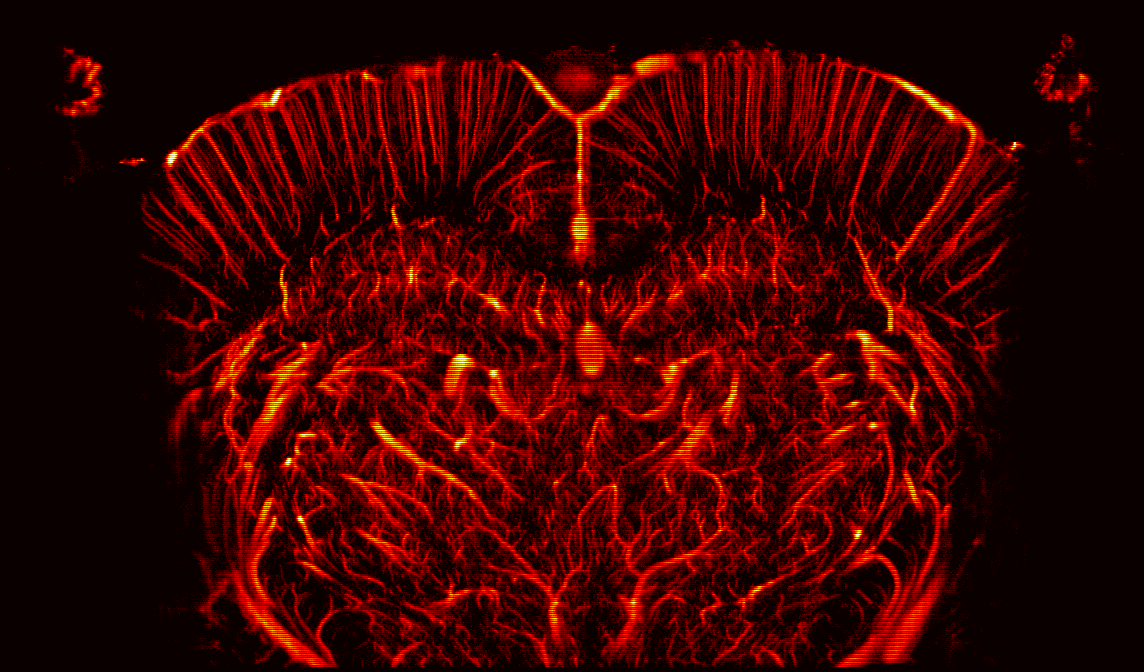}
        };
        \spy [closeup,magnification=3] on ($(FigA)+(-0.305,+0.075)$) 
        in node[largewindow,anchor=north east] at ($(FigA.north west) - (0,0.02)$);
        \spy [closeup,magnification=2] on ($(FigA)+(+0.045,+0.02)$) 
        in node[largewindow,anchor=east]       at ($(FigA.west)$);
        \spy [closeup,magnification=2] on ($(FigA)+(-0.265,-0.075)$) 
        in node[largewindow,anchor=south east] at ($(FigA.south west) + (0,0.02)$);
        \end{tikzpicture}
    }
    \overlayRectangle{7.45cm}{1.25cm}{.375cm}{.5cm}
    \subcaption{U-Net~\cite{van2020super} + NMS from B-mode frames \label{fig:rat18:unet}}
    \end{minipage}
    \begin{minipage}[b]{.49\linewidth}
    \resizebox{\textwidth}{!}{
        \begin{tikzpicture}[x=6cm, y=6cm, spy using outlines={every spy on node/.append style={smallwindow}}]
        \node[anchor=south] (FigA) at (0,0) {\includegraphics[height=3cm,trim=9cm 17.5cm 18cm 2.5cm,clip]{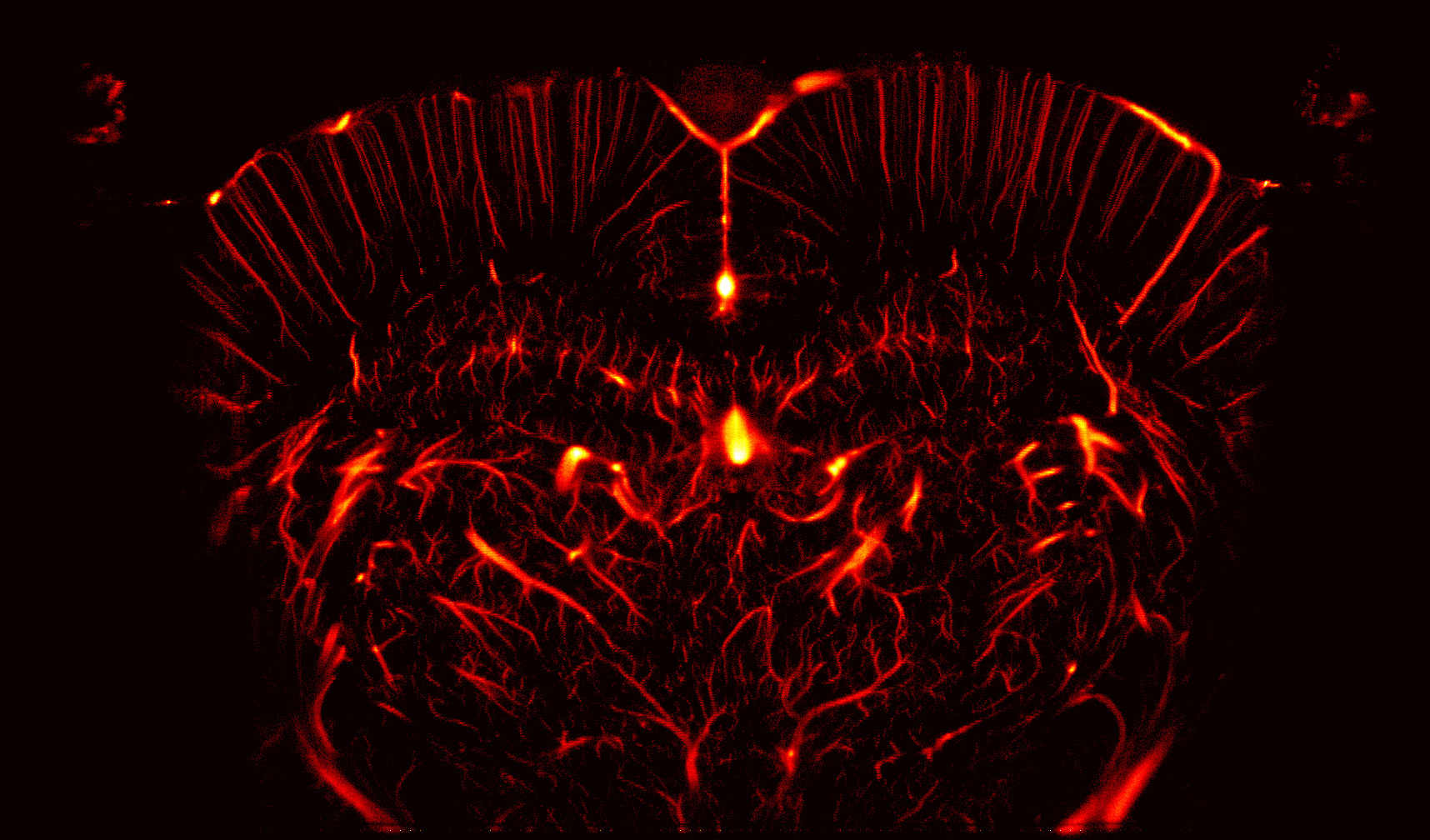}};
        \spy [closeup,magnification=3] on ($(FigA)+(-0.305,+0.075)$) 
        in node[largewindow,anchor=north east] at ($(FigA.north west) - (0,0.02)$);
        \spy [closeup,magnification=2] on ($(FigA)+(+0.045,+0.02)$) 
        in node[largewindow,anchor=east]       at ($(FigA.west)$);
        \spy [closeup,magnification=2] on ($(FigA)+(-0.265,-0.075)$) 
        in node[largewindow,anchor=south east] at ($(FigA.south west) + (0,0.02)$);
        \end{tikzpicture}
    }
    \overlayRectangle{7.45cm}{1.25cm}{.375cm}{.5cm}
    \subcaption{Radial Symmetry~\cite{heiles2022pala} from B-mode frames\label{fig:rat18:rs}}
    \end{minipage}
    \begin{minipage}[b]{.49\linewidth}
    \resizebox{\textwidth}{!}{
        \begin{tikzpicture}[x=6cm, y=6cm, spy using outlines={every spy on node/.append style={smallwindow}}]
        \node[anchor=south] (FigA) at (0,0) {\includegraphics[height=3cm,trim=9cm 17.5cm 18cm 2.5cm,clip]{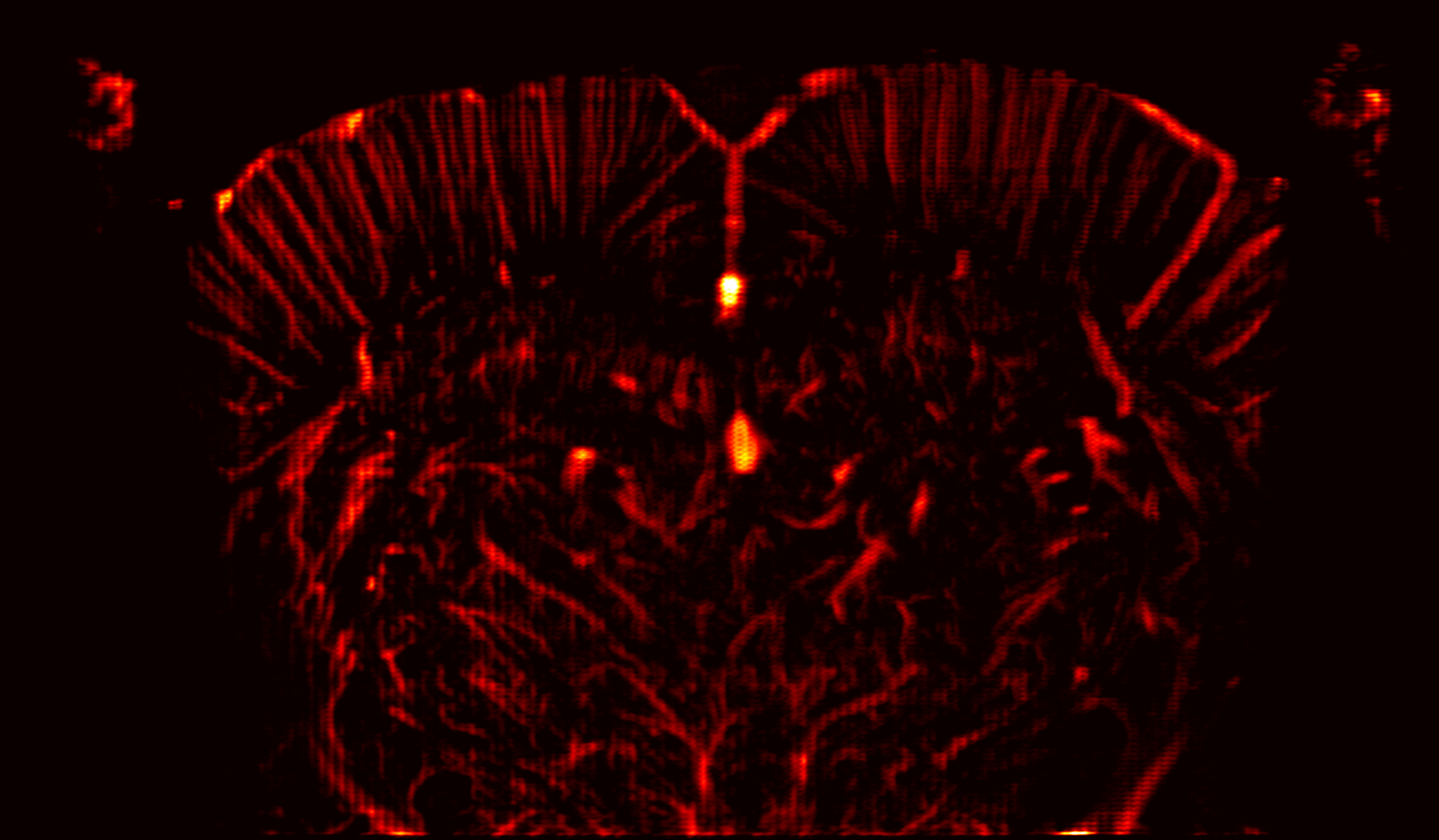}};
        \spy [closeup,magnification=3] on ($(FigA)+(-0.305,+0.075)$) 
        in node[largewindow,anchor=north east] at ($(FigA.north west) - (0,0.02)$);
        \spy [closeup,magnification=2] on ($(FigA)+(+0.045,+0.02)$) 
        in node[largewindow,anchor=east]       at ($(FigA.west)$);
        \spy [closeup,magnification=2] on ($(FigA)+(-0.265,-0.075)$) 
        in node[largewindow,anchor=south east] at ($(FigA.south west) + (0,0.02)$);
        \end{tikzpicture}
    }
    \overlayRectangle{7.45cm}{1.25cm}{.375cm}{.5cm}
    \subcaption{SG-SPCN from B-mode frames without NMS \label{fig:rat18:sgspcnbmodewonms}}
    \end{minipage}
    \centering
    \begin{minipage}[b]{.49\linewidth}
    \resizebox{\textwidth}{!}{
        \begin{tikzpicture}[x=6cm, y=6cm, spy using outlines={every spy on node/.append style={smallwindow}}]
        \node[anchor=south] (FigA) at (1.0,0) {\includegraphics[height=3cm,trim=9cm 17.5cm 18cm 2.5cm,clip]{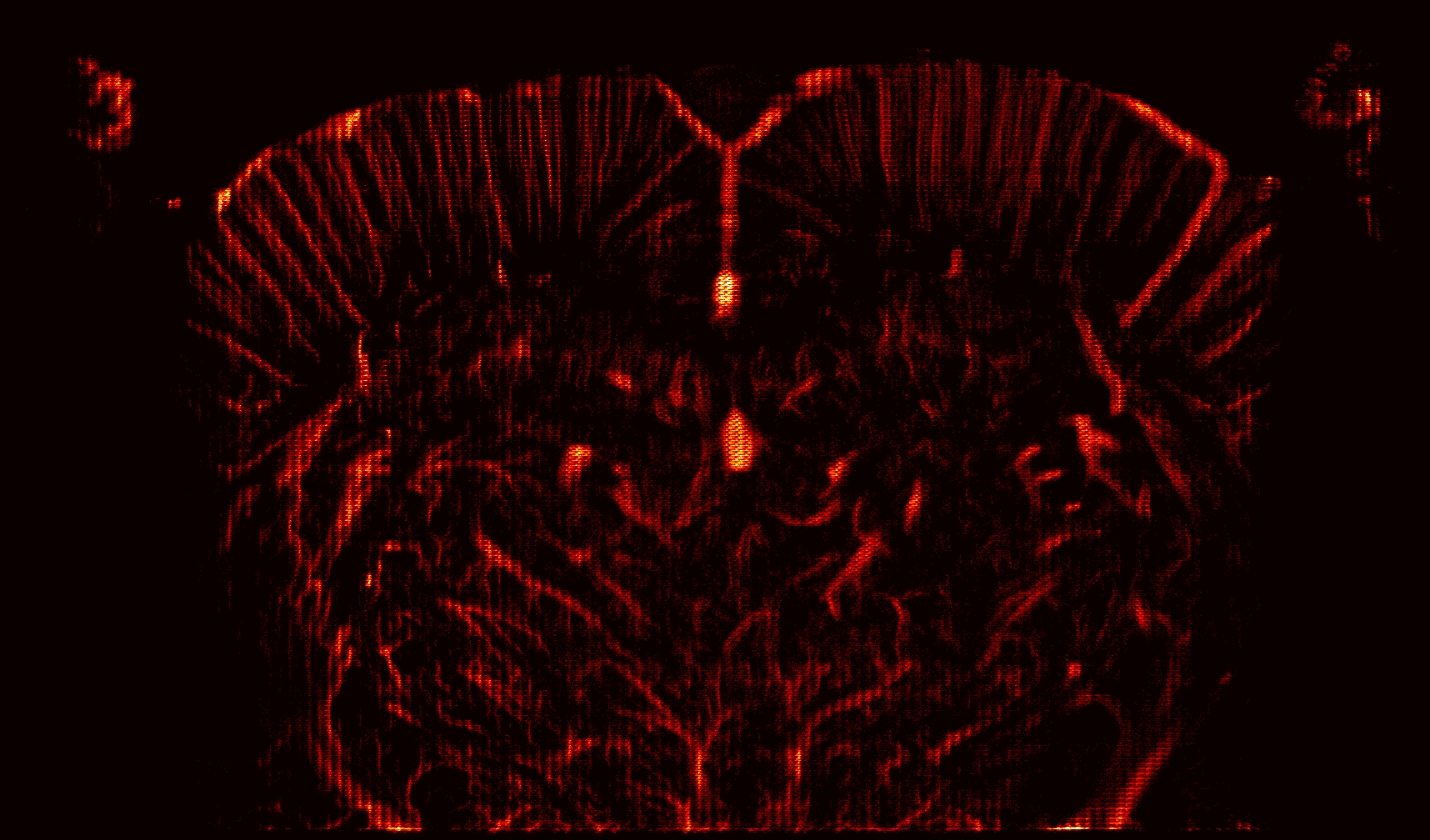}};
        \spy [closeup,magnification=3] on ($(FigA)+(-0.305,+0.075)$) 
        in node[largewindow,anchor=north east] at ($(FigA.north west) - (0,0.02)$);
        \spy [closeup,magnification=2] on ($(FigA)+(+0.045,+0.02)$) 
        in node[largewindow,anchor=east]       at ($(FigA.west)$);
        \spy [closeup,magnification=2] on ($(FigA)+(-0.265,-0.075)$) 
        in node[largewindow,anchor=south east] at ($(FigA.south west) + (0,0.02)$);
        \end{tikzpicture}
    }
    \overlayRectangle{7.45cm}{1.25cm}{.375cm}{.5cm}
    \subcaption{SG-SPCN + NMS from B-mode frames \label{fig:rat18:sgspcnbmode}}
    \end{minipage}
    \begin{minipage}[b]{.49\linewidth}
    \resizebox{\textwidth}{!}{
        \begin{tikzpicture}[x=6cm, y=6cm, spy using outlines={every spy on node/.append style={smallwindow}}]
        \node[anchor=south] (FigA) at (1.0,0) {\includegraphics[height=3cm,trim=9cm 17.5cm 18cm 2.5cm,clip]{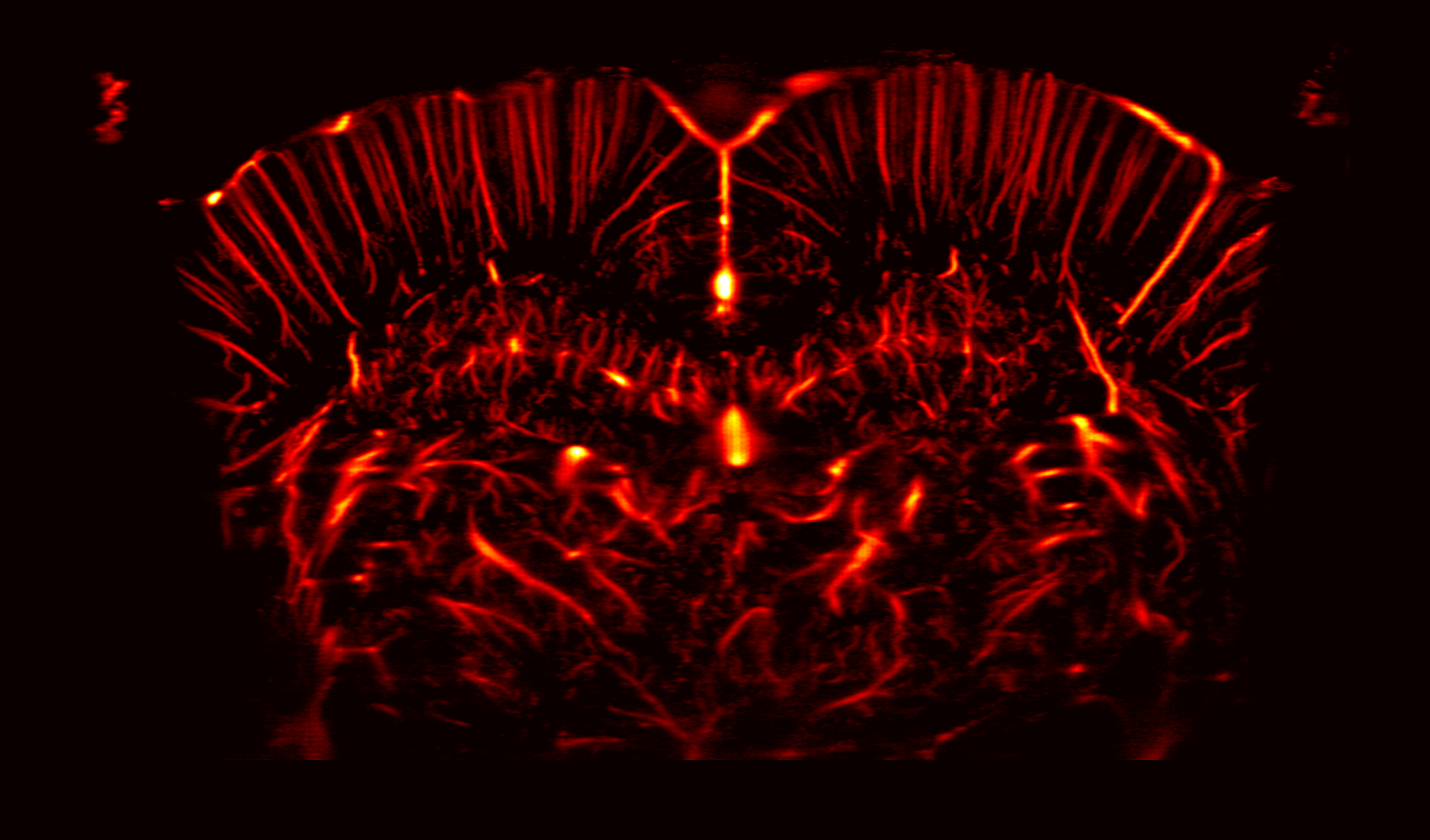}};
        \spy [closeup,magnification=3] on ($(FigA)+(-0.305,+0.075)$) 
        in node[largewindow,anchor=north east] at ($(FigA.north west) - (0,0.02)$);
        \spy [closeup,magnification=2] on ($(FigA)+(+0.045,+0.02)$) 
        in node[largewindow,anchor=east]       at ($(FigA.west)$);
        \spy [closeup,magnification=2] on ($(FigA)+(-0.265,-0.075)$) 
        in node[largewindow,anchor=south east] at ($(FigA.south west) + (0,0.02)$);
        \end{tikzpicture}
    }
    \overlayRectangle{7.45cm}{1.25cm}{.375cm}{.5cm}
    \subcaption{SG-SPCN from RF$\to$I/Q frames without NMS\label{fig:rat18:sgspcnrfwonms}}
    \end{minipage}
    \centering
    \begin{minipage}[b]{.49\linewidth}
    \resizebox{\textwidth}{!}{
        \begin{tikzpicture}[x=6cm, y=6cm, spy using outlines={every spy on node/.append style={smallwindow}}]
        \node[anchor=south] (FigA) at (1.0,0) {\includegraphics[height=3cm,trim=9cm 17.5cm 18cm 2.5cm,clip]{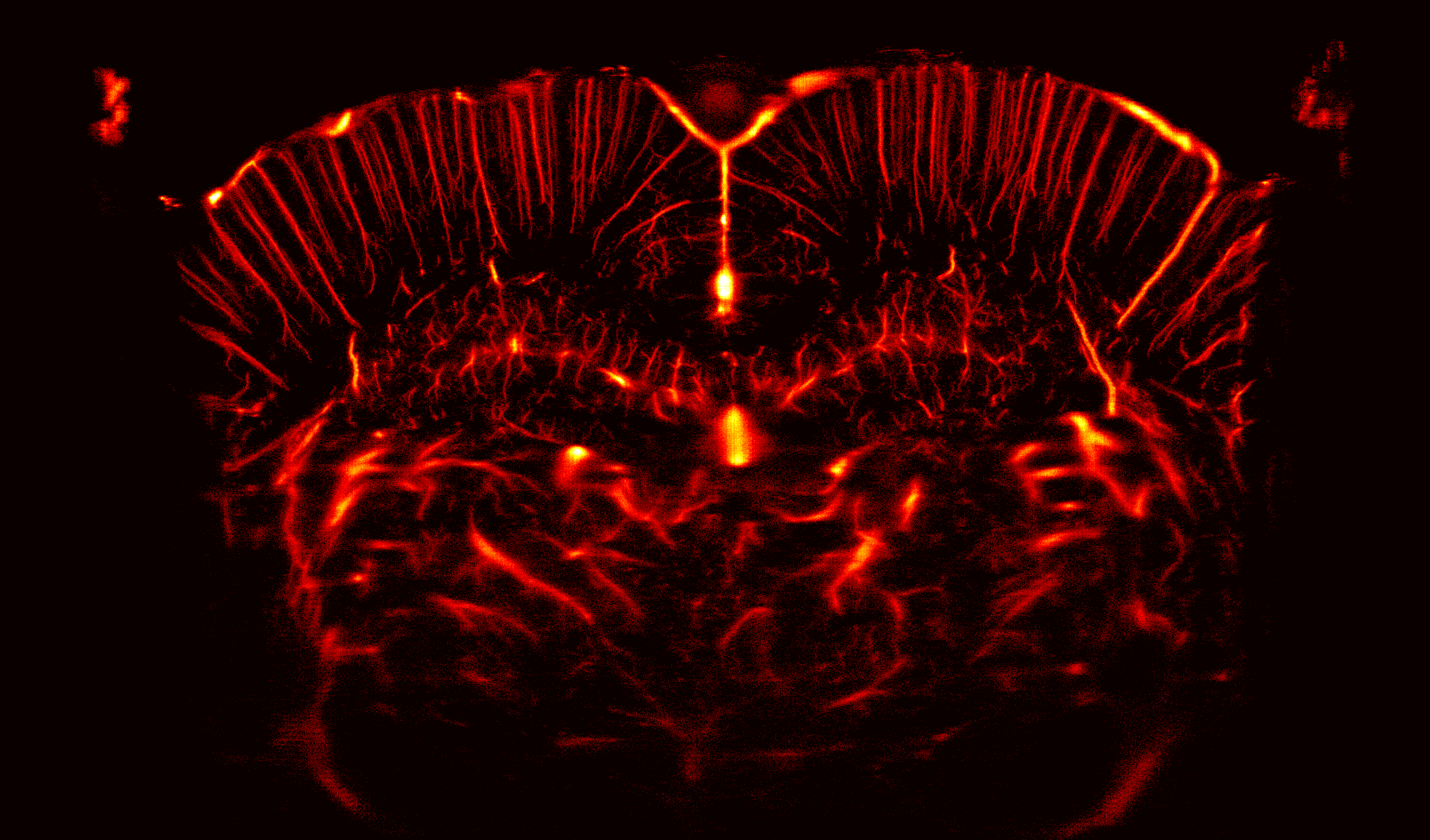}};
        \spy [closeup,magnification=3] on ($(FigA)+(-0.305,+0.075)$) 
        in node[largewindow,anchor=north east] at ($(FigA.north west) - (0,0.02)$);
        \spy [closeup,magnification=2] on ($(FigA)+(+0.045,+0.02)$) 
        in node[largewindow,anchor=east]       at ($(FigA.west)$);
        \spy [closeup,magnification=2] on ($(FigA)+(-0.265,-0.075)$) 
        in node[largewindow,anchor=south east] at ($(FigA.south west) + (0,0.02)$);
        \end{tikzpicture}
    }
    \overlayRectangle{7.45cm}{1.25cm}{.375cm}{.5cm}
    \subcaption{SG-SPCN + NMS from RF$\to$I/Q frames \label{fig:rat18:sgspcnrf}}
    \end{minipage}
    \caption{\textbf{In vivo ULM results} that show the microvascular structure of rat brains. The images depict accumulated localizations from rat-20 processed with 5 compounded plane waves in (\subref{fig:rat20:unet})~to~(\subref{fig:rat20:sgspcnrf}) and from rat-18 with 3 compounded plane waves in (\subref{fig:rat18:unet})~to~(\subref{fig:rat18:sgspcnrf}) using 128 transducer channel data of $120\times800$ frames each. 
    Blue rectangles highlight magnified views for better comparison. To illustrate the impact of NMS, (\subref{fig:rat18:sgspcnbmodewonms}) and (\subref{fig:rat18:sgspcnrfwonms}) show the mean of predictions $f(\mathbf{X})$ in the absence of NMS. 
    }
    \label{fig:rat18}
\end{figure*}


\newcommand{\scalebarbackground}[5][white]{
 \begin{tikzpicture}
  \node[anchor=south west,inner sep=0] (image) { #2 };
  \begin{scope}[x={(image.south east)},y={(image.north west)}]
   \fill [fill=mblue, fill opacity=0.5] (0.025,1.5em) rectangle (#5*#4/#3+0.025,0.3em);
   \draw [#1, line width=0.2em] (0.025,1.5em) -- node[below,inner sep=0.225em, font=\footnotesize] {\SI{#5}{\milli \meter}} (#5*#4/#3+0.025,1.5em);
  \end{scope}
 \end{tikzpicture}
}

\begin{figure}[!t]
    \begin{minipage}{\linewidth}
    \begin{minipage}{0.59\linewidth}
        \centering
        \scalebarbackground{
        \includegraphics[width=.81\linewidth]{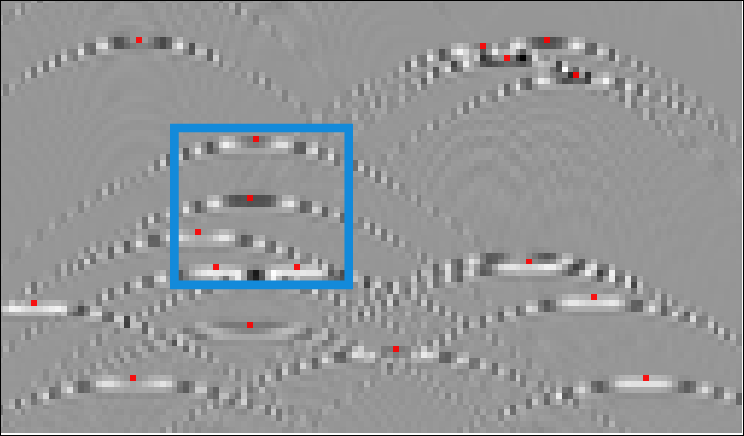}
        }{12.8}{1}{3}
        \subcaption{\textit{In silico} RF$\to$I/Q test frame\label{fig:sub:rf_insilico_lores}}
    \end{minipage}
    \hfill
    \begin{minipage}{0.39\linewidth}
        \centering
        \includegraphics[width=.8\linewidth]{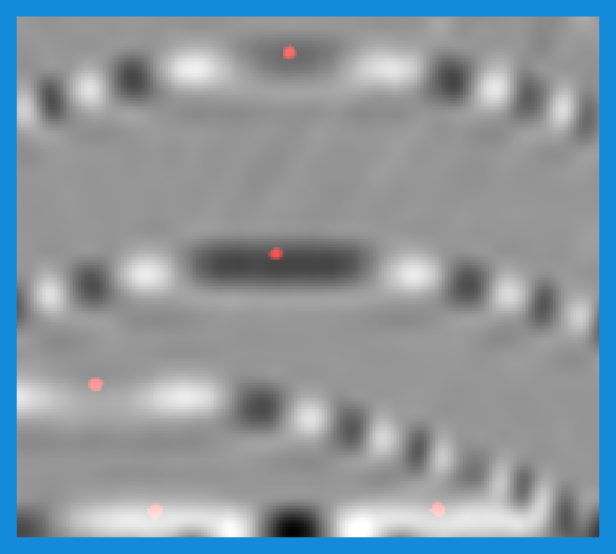}
        \subcaption{Magnification of (\subref{fig:sub:rf_insilico_lores}) \label{fig:sub:rf_insilico_hires}}
    \end{minipage}
    \end{minipage}
    \vspace{.1cm}
    \vfill
    \begin{minipage}{\linewidth}
    \begin{minipage}{0.59\linewidth}
        \centering
        \scalebarbackground{
        \includegraphics[width=.81\linewidth]{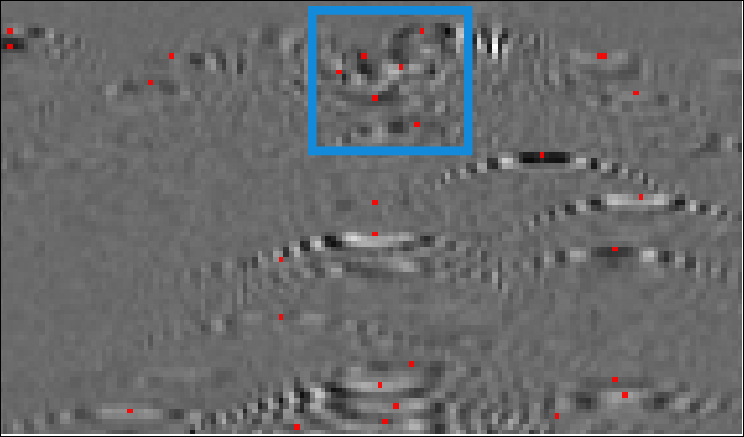}
        }{12.8}{1}{3}
        \subcaption{\textit{In vivo} RF$\to$I/Q frame\label{fig:sub:rf_invivo_lores}}
    \end{minipage}
    \hfill
    \begin{minipage}{0.39\linewidth}
        \centering
        \includegraphics[width=.8\linewidth]{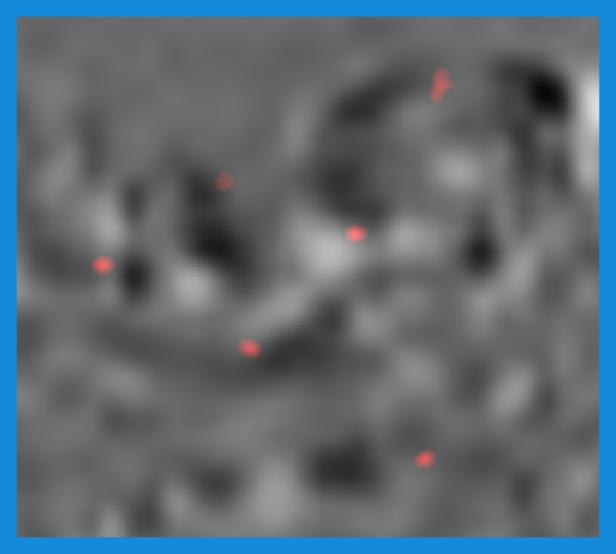}
        \subcaption{Magnification of (\subref{fig:sub:rf_invivo_lores})\label{fig:sub:rf_invivo_hires}}
    \end{minipage}
    \end{minipage}
    \vspace{.1cm}
    \vfill
    \begin{minipage}{\linewidth}
    \begin{minipage}{0.59\linewidth}
        \centering
        \scalebarbackground{
        \includegraphics[width=.81\linewidth]{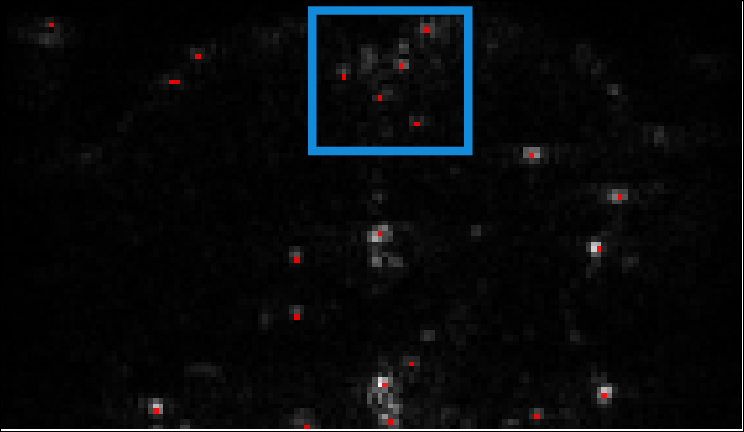}
        }{14.3}{0.985}{3}%
        \subcaption{\textit{In vivo} B-mode frame\label{fig:sub:bmode_lores}}
    \end{minipage}
    \hfill
    \begin{minipage}{0.39\linewidth}
        \centering
        \includegraphics[width=.8\linewidth]{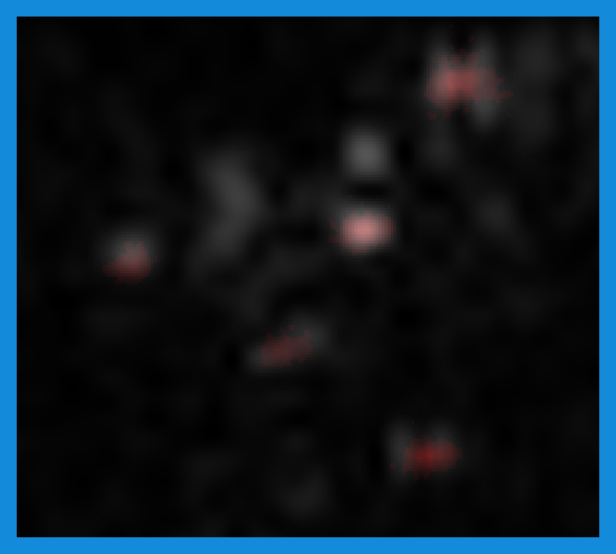}
        \subcaption{Magnification of (\subref{fig:sub:bmode_lores})\label{fig:sub:bmode_hires}}
    \end{minipage}
    \end{minipage}
    \caption{\textbf{\mbox{SG-SPCN} localizations} overlaid as red blobs on synthetic and real frames. Blue rectangles show the magnified region. The images in (\subref{fig:sub:rf_insilico_lores})~to~(\subref{fig:sub:rf_invivo_hires}) depict the I component of I/Q input signals and (\subref{fig:sub:rf_invivo_lores})~to~(\subref{fig:sub:bmode_hires}) are from rat-18 (frame 2400) at $R=12$. The results show our method's ability to effectively transfer knowledge from synthetic to unseen \textit{in vivo} data. %
    }
    \label{fig:mb_localization}
\end{figure}

Another key observation in our study is the ability of our network trained on synthetic RF data to generalize effectively when applied to real data. This suggests that the knowledge about wavefronts gained from synthetic data suffices to bridge the domain gap, contributing to improved performance in practical applications. In support of this, we notice that mSPCN performs well for \textit{in vivo} RF channel data in our prior study~\cite{hahne:23:ius}, but falls short in achieving the same from the B-mode inputs even when provided a semi-global contraction block (see Fig.~\ref{fig:rat18}). 
%
To further scrutinize the network's ability for domain adaptation, we compile localization results for an intermediate frame in Fig.~\ref{fig:mb_localization} and overlay $f(\mathbf{X})$ predictions in red color. Comparing these examples, one may note the precise channel data localizations 
while B-mode predictions expose larger distributions.

Temporal filtering is a crucial component of ULM rendering. Our real data results highlight the feasibility of employing temporal filters within RF channels without compromising on image quality. This finding aligns with related research in non-destructive testing~\cite{rao2022ultrasonic} and Doppler imaging~\cite{Pialot:2023} where filtering prior to beamforming proves to be effective. 

\section{Conclusion}
\label{sec:conclusion}

This study provides valuable insights into the role of beamforming and RF data for ULM. Our proposed method demonstrates the feasibility of localizing microbubbles using \textit{in vivo} data without relying on \mbox{delay-and-sum} beamforming. We achieve this through the innovative use of a \mbox{super-resolution} deep neural network and non-maximum suppression to identify distinct center coordinates of incoming wavefronts with remarkable precision. To enable \mbox{RF-ULM} rendering, we combine this network with custom forward and backward transformations to map points between RF and \mbox{B-mode} coordinate spaces. 
%
Our extensive benchmark study reveals that omitting beamforming in ULM not only reduces the time complexity, which holds particular promise for 3-D scenarios, but also enhances the mean localization accuracy by more than 20\%. 
We demonstrate that our network, trained on synthetic RF data, exhibits effective generalization when applied to real data. This highlights the significance of the knowledge acquired from synthetic data in addressing the domain gap.
These findings hold promise for advancing future ULM pipelines, potentially contributing to the clinical adoption of this groundbreaking technology.
Acknowledging the importance of our study's findings, which rely on high frame rates and linear scatting, we recognize that extending our findings to other applications exceeds our current technical feasibility and benchmark comparison. Therefore, we emphasize the need for further experimental validation before extrapolating our results to broader contexts.

\section*{Acknowledgment}
This study is funded in part by the Hasler Foundation under number 22027, and the authors wish to express their appreciation to the foundation for their support.

\bibliographystyle{IEEEtran}
\bibliography{main}

\begin{thebibliography}{10}
\providecommand{\url}[1]{#1}
\csname url@samestyle\endcsname
\providecommand{\newblock}{\relax}
\providecommand{\bibinfo}[2]{#2}
\providecommand{\BIBentrySTDinterwordspacing}{\spaceskip=0pt\relax}
\providecommand{\BIBentryALTinterwordstretchfactor}{4}
\providecommand{\BIBentryALTinterwordspacing}{\spaceskip=\fontdimen2\font plus
\BIBentryALTinterwordstretchfactor\fontdimen3\font minus
  \fontdimen4\font\relax}
\providecommand{\BIBforeignlanguage}[2]{{%
\expandafter\ifx\csname l@#1\endcsname\relax
\typeout{** WARNING: IEEEtran.bst: No hyphenation pattern has been}%
\typeout{** loaded for the language `#1'. Using the pattern for}%
\typeout{** the default language instead.}%
\else
\language=\csname l@#1\endcsname
\fi
#2}}
\providecommand{\BIBdecl}{\relax}
\BIBdecl

\bibitem{heiles2022pala}
B.~Heiles, A.~Chavignon, V.~Hingot, P.~Lopez, E.~Teston, and O.~Couture,
  ``Performance benchmarking of microbubble-localization algorithms for
  ultrasound localization microscopy,'' \emph{Nature Biomedical Engineering},
  vol.~6, no.~5, pp. 605--616, 2022.

\bibitem{errico2015ultrafast}
C.~Errico, J.~Pierre, S.~Pezet, Y.~Desailly, Z.~Lenkei, O.~Couture, and
  M.~Tanter, ``Ultrafast ultrasound localization microscopy for deep
  super-resolution vascular imaging,'' \emph{Nature}, vol. 527, no. 7579, pp.
  499--502, 2015.

\bibitem{StateOfTheArt}
O.~Couture, V.~Hingot, B.~Heiles, P.~Muleki-Seya, and M.~Tanter, ``Ultrasound
  localization microscopy and super-resolution: A state of the art,''
  \emph{IEEE Transactions on Ultrasonics, Ferroelectrics, and Frequency
  Control}, vol.~65, no.~8, pp. 1304--1320, 2018.

\bibitem{song:2018}
P.~Song, A.~Manduca, J.~Trzasko, R.~Daigle, and S.~Chen, ``On the effects of
  spatial sampling quantization in super-resolution ultrasound microvessel
  imaging,'' \emph{IEEE transactions on ultrasonics, ferroelectrics, and
  frequency control}, vol.~65, no.~12, pp. 2264--2276, 2018.

\bibitem{van2020super}
R.~J. van Sloun, O.~Solomon, M.~Bruce, Z.~Z. Khaing, H.~Wijkstra, Y.~C. Eldar,
  and M.~Mischi, ``Super-resolution ultrasound localization microscopy through
  deep learning,'' \emph{IEEE transactions on medical imaging}, vol.~40, no.~3,
  pp. 829--839, 2020.

\bibitem{liu2020deep}
X.~Liu, T.~Zhou, M.~Lu, Y.~Yang, Q.~He, and J.~Luo, ``Deep learning for
  ultrasound localization microscopy,'' \emph{IEEE transactions on medical
  imaging}, vol.~39, no.~10, pp. 3064--3078, 2020.

\bibitem{gulm:2023}
C.~Hahne and R.~Sznitman, ``Geometric ultrasound localization microscopy,'' in
  \emph{Medical Image Computing and Computer Assisted Intervention--MICCAI
  2023: 26th International Conference, Vancouver, Canada, October 8--12, 2023,
  Proceedings, Part VII 26}.\hskip 1em plus 0.5em minus 0.4em\relax Springer,
  2023, pp. 1--10.

\bibitem{chabouh2021spherical}
G.~Chabouh, B.~Dollet, C.~Quilliet, and G.~Coupier, ``Spherical oscillations of
  encapsulated microbubbles: Effect of shell compressibility and anisotropy,''
  \emph{The Journal of the Acoustical Society of America}, vol. 149, no.~2, pp.
  1240--1257, 2021.

\bibitem{chabouh:2023:buckling}
G.~Chabouh, B.~van Elburg, M.~Versluis, T.~Segers, C.~Quilliet, and G.~Coupier,
  ``Buckling of lipidic ultrasound contrast agents under quasi-static load,''
  \emph{Philosophical Transactions of the Royal Society A: Mathematical,
  Physical and Engineering Sciences}, vol. 381, no. 2244, p. 20220025, 2023.

\bibitem{bar:2021}
O.~Bar-Shira, A.~Grubstein, Y.~Rapson, D.~Suhami, E.~Atar, K.~Peri-Hanania,
  R.~Rosen, and Y.~C. Eldar, ``Learned super resolution ultrasound for improved
  breast lesion characterization,'' in \emph{Medical Image Computing and
  Computer Assisted Intervention--MICCAI 2021: 24th International Conference,
  Strasbourg, France, September 27--October 1, 2021, Proceedings, Part VII
  24}.\hskip 1em plus 0.5em minus 0.4em\relax Springer, 2021, pp. 109--118.

\bibitem{demene2021transcranial}
C.~Demen{\'e}, J.~Robin, A.~Dizeux, B.~Heiles, M.~Pernot, M.~Tanter, and
  F.~Perren, ``Transcranial ultrafast ultrasound localization microscopy of
  brain vasculature in patients,'' \emph{Nature biomedical engineering},
  vol.~5, no.~3, pp. 219--228, 2021.

\bibitem{bodard2023ultrasound}
S.~Bodard, L.~Denis, V.~Hingot, A.~Chavignon, O.~H{\'e}l{\'e}non,
  D.~Anglicheau, O.~Couture, and J.-M. Correas, ``Ultrasound localization
  microscopy of the human kidney allograft on a clinical ultrasound scanner,''
  \emph{Kidney International}, vol. 103, no.~5, pp. 930--935, 2023.

\bibitem{song2023super}
P.~Song, J.~M. Rubin, and M.~R. Lowerison, ``Super-resolution ultrasound
  microvascular imaging: Is it ready for clinical use?'' \emph{Zeitschrift
  f{\"u}r Medizinische Physik}, 2023.

\bibitem{chavignon:2021}
A.~Chavignon, B.~Heiles, V.~Hingot, C.~Orset, D.~Vivien, and O.~Couture, ``3d
  transcranial ultrasound localization microscopy in the rat brain with a
  multiplexed matrix probe,'' \emph{IEEE Transactions on Biomedical
  Engineering}, vol.~69, no.~7, pp. 2132--2142, 2021.

\bibitem{heiles2022volumetric}
B.~Heiles, A.~Chavignon, A.~Bergel, V.~Hingot, H.~Serroune, D.~Maresca,
  S.~Pezet, M.~Pernot, M.~Tanter, and O.~Couture, ``Volumetric ultrasound
  localization microscopy of the whole rat brain microvasculature,'' \emph{IEEE
  Open Journal of Ultrasonics, Ferroelectrics, and Frequency Control}, vol.~2,
  pp. 261--282, 2022.

\bibitem{demeulenaere2022coronary}
O.~Demeulenaere, Z.~Sandoval, P.~Mateo, A.~Dizeux, O.~Villemain, R.~Gallet,
  B.~Ghaleh, T.~Deffieux, C.~Dem{\'e}n{\'e}, M.~Tanter \emph{et~al.},
  ``Coronary flow assessment using 3-dimensional ultrafast ultrasound
  localization microscopy,'' \emph{Cardiovascular Imaging}, vol.~15, no.~7, pp.
  1193--1208, 2022.

\bibitem{chavignon20223d}
A.~Chavignon, V.~Hingot, C.~Orset, D.~Vivien, and O.~Couture, ``3d transcranial
  ultrasound localization microscopy for discrimination between ischemic and
  hemorrhagic stroke in early phase,'' \emph{Scientific Reports}, vol.~12,
  no.~1, p. 14607, 2022.

\bibitem{yan2022super}
J.~Yan, T.~Zhang, J.~Broughton-Venner, P.~Huang, and M.-X. Tang,
  ``Super-resolution ultrasound through sparsity-based deconvolution and
  multi-feature tracking,'' \emph{IEEE Transactions on Medical Imaging},
  vol.~41, no.~8, pp. 1938--1947, 2022.

\bibitem{you2022curvelet}
Q.~You, J.~D. Trzasko, M.~R. Lowerison, X.~Chen, Z.~Dong, N.~V. ChandraSekaran,
  D.~A. Llano, S.~Chen, and P.~Song, ``Curvelet transform-based sparsity
  promoting algorithm for fast ultrasound localization microscopy,'' \emph{IEEE
  transactions on medical imaging}, vol.~41, no.~9, pp. 2385--2398, 2022.

\bibitem{hahne:23:ius}
C.~Hahne, G.~Chabouh, O.~Couture, and R.~Sznitman, ``Learning super-resolution
  ultrasound localization microscopy from radio-frequency data,'' in \emph{2023
  IEEE International Ultrasonics Symposium (IUS)}, 2023, pp. 1--4.

\bibitem{wang2022general}
R.~Wang and W.-N. Lee, ``A general deep learning model for ultrasound
  localization microscopy,'' in \emph{2022 IEEE International Ultrasonics
  Symposium (IUS)}.\hskip 1em plus 0.5em minus 0.4em\relax IEEE, 2022, pp.
  1--4.

\bibitem{long2022super}
F.~Long and W.~Zhang, ``Super resolution ultrasound imaging using deep learning
  based micro-bubbles localization,'' in \emph{2022 IEEE International
  Ultrasonics Symposium (IUS)}.\hskip 1em plus 0.5em minus 0.4em\relax IEEE,
  2022, pp. 1--5.

\bibitem{sui2022generative}
Y.~Sui, X.~Guo, J.~Yu, D.~Ta, and K.~Xu, ``Generative adversarial nets for
  ultrafast ultrasound localization microscopy reconstruction,'' in \emph{2022
  IEEE International Ultrasonics Symposium (IUS)}.\hskip 1em plus 0.5em minus
  0.4em\relax IEEE, 2022, pp. 1--4.

\bibitem{gharamaleki2022transformer}
S.~K. Gharamaleki, B.~Helfield, and H.~Rivaz, ``Transformer-based microbubble
  localization,'' in \emph{2022 IEEE International Ultrasonics Symposium
  (IUS)}.\hskip 1em plus 0.5em minus 0.4em\relax IEEE, 2022, pp. 1--4.

\bibitem{liu2022ultrasound}
X.~Liu and M.~Almekkawy, ``Ultrasound super resolution using vision transformer
  with convolution projection operation,'' in \emph{2022 IEEE International
  Ultrasonics Symposium (IUS)}.\hskip 1em plus 0.5em minus 0.4em\relax IEEE,
  2022, pp. 1--4.

\bibitem{milecki2021deep}
L.~Milecki, J.~Por{\'e}e, H.~Belgharbi, C.~Bourquin, R.~Damseh,
  P.~Delafontaine-Martel, F.~Lesage, M.~Gasse, and J.~Provost, ``A deep
  learning framework for spatiotemporal ultrasound localization microscopy,''
  \emph{IEEE Transactions on Medical Imaging}, vol.~40, no.~5, pp. 1428--1437,
  2021.

\bibitem{chen2022deep}
X.~Chen, M.~R. Lowerison, Z.~Dong, A.~Han, and P.~Song, ``Deep learning-based
  microbubble localization for ultrasound localization microscopy,'' \emph{IEEE
  transactions on ultrasonics, ferroelectrics, and frequency control}, vol.~69,
  no.~4, pp. 1312--1325, 2022.

\bibitem{chen2023deep}
X.~Chen, M.~R. Lowerison, Z.~Dong, N.~V. Chandra~Sekaran, D.~A. Llano, and
  P.~Song, ``Localization free super-resolution microbubble velocimetry using a
  long short-term memory neural network,'' \emph{IEEE Transactions on Medical
  Imaging}, vol.~42, no.~8, pp. 2374--2385, 2023.

\bibitem{nair2018deep}
A.~A. Nair, T.~D. Tran, A.~Reiter, and M.~A.~L. Bell, ``A deep learning based
  alternative to beamforming ultrasound images,'' in \emph{2018 IEEE
  International conference on acoustics, speech and signal processing
  (ICASSP)}.\hskip 1em plus 0.5em minus 0.4em\relax IEEE, 2018, pp. 3359--3363.

\bibitem{zhang2021ultrasound}
J.~Zhang, Q.~He, Y.~Xiao, H.~Zheng, C.~Wang, and J.~Luo, ``Ultrasound image
  reconstruction from plane wave radio-frequency data by self-supervised deep
  neural network,'' \emph{Medical Image Analysis}, vol.~70, p. 102018, 2021.

\bibitem{couture2011microbubble}
O.~Couture, B.~Besson, G.~Montaldo, M.~Fink, and M.~Tanter, ``Microbubble
  ultrasound super-localization imaging (musli),'' in \emph{2011 IEEE
  International Ultrasonics Symposium}.\hskip 1em plus 0.5em minus 0.4em\relax
  IEEE, 2011, pp. 1285--1287.

\bibitem{corazza2022beamforming}
A.~Corazza, P.~Muleki-Seya, A.~W. Aissani, O.~Couture, A.~Basarab, and
  B.~Nicolas, ``Microbubble detection with adaptive beamforming for ultrasound
  localization microscopy,'' in \emph{2022 IEEE International Ultrasonics
  Symposium (IUS)}.\hskip 1em plus 0.5em minus 0.4em\relax IEEE, 2022, pp.
  1--4.

\bibitem{youn2020detection}
J.~Youn, M.~L. Ommen, M.~B. Stuart, E.~V. Thomsen, N.~B. Larsen, and J.~A.
  Jensen, ``Detection and localization of ultrasound scatterers using
  convolutional neural networks,'' \emph{IEEE Transactions on Medical Imaging},
  vol.~39, no.~12, pp. 3855--3867, 2020.

\bibitem{blanken2022sr}
N.~Blanken, J.~M. Wolterink, H.~Delingette, C.~Brune, M.~Versluis, and
  G.~Lajoinie, ``Super-resolved microbubble localization in single-channel
  ultrasound rf signals using deep learning,'' \emph{IEEE Transactions on
  Medical Imaging}, vol.~41, no.~9, pp. 2532--2542, 2022.

\bibitem{ronneberger2015u}
O.~Ronneberger, P.~Fischer, and T.~Brox, ``U-net: Convolutional networks for
  biomedical image segmentation,'' in \emph{Medical Image Computing and
  Computer-Assisted Intervention--MICCAI 2015: 18th International Conference,
  Munich, Germany, October 5-9, 2015, Proceedings, Part III 18}.\hskip 1em plus
  0.5em minus 0.4em\relax Springer, 2015, pp. 234--241.

\bibitem{shi2016real}
W.~Shi, J.~Caballero, F.~Husz{\'a}r, J.~Totz, A.~P. Aitken, R.~Bishop,
  D.~Rueckert, and Z.~Wang, ``Real-time single image and video super-resolution
  using an efficient sub-pixel convolutional neural network,'' in
  \emph{Proceedings of the IEEE conference on computer vision and pattern
  recognition}, 2016, pp. 1874--1883.

\bibitem{lim2017enhanced}
B.~Lim, S.~Son, H.~Kim, S.~Nah, and K.~Mu~Lee, ``Enhanced deep residual
  networks for single image super-resolution,'' in \emph{Proceedings of the
  IEEE conference on computer vision and pattern recognition workshops}, 2017,
  pp. 136--144.

\bibitem{li2019cr}
H.~Li, J.~Fang, S.~Liu, X.~Liang, X.~Yang, Z.~Mai, M.~T. Van, T.~Wang, Z.~Chen,
  and D.~Ni, ``Cr-unet: A composite network for ovary and follicle segmentation
  in ultrasound images,'' \emph{IEEE journal of biomedical and health
  informatics}, vol.~24, no.~4, pp. 974--983, 2019.

\bibitem{neubeck2006nms}
A.~Neubeck and L.~Van~Gool, ``Efficient non-maximum suppression,'' in
  \emph{18th International Conference on Pattern Recognition (ICPR'06)},
  vol.~3, 2006, pp. 850--855.

\bibitem{kubat1997addressing}
M.~Kubat, S.~Matwin \emph{et~al.}, ``Addressing the curse of imbalanced
  training sets: one-sided selection,'' in \emph{Icml}, vol.~97, no.~1.\hskip
  1em plus 0.5em minus 0.4em\relax Citeseer, 1997, p. 179.

\bibitem{ester1996density}
M.~Ester, H.-P. Kriegel, J.~Sander, X.~Xu \emph{et~al.}, ``A density-based
  algorithm for discovering clusters in large spatial databases with noise,''
  in \emph{kdd}, vol.~96, no.~34, 1996, pp. 226--231.

\bibitem{demene2015spatiotemporal}
C.~Demené, T.~Deffieux, M.~Pernot, B.-F. Osmanski, V.~Biran, J.-L. Gennisson,
  L.-A. Sieu, A.~Bergel, S.~Franqui, J.-M. Correas, I.~Cohen, O.~Baud, and
  M.~Tanter, ``Spatiotemporal clutter filtering of ultrafast ultrasound data
  highly increases doppler and fultrasound sensitivity,'' \emph{IEEE
  Transactions on Medical Imaging}, vol.~34, no.~11, pp. 2271--2285, 2015.

\bibitem{baranger2018adaptive}
J.~Baranger, B.~Arnal, F.~Perren, O.~Baud, M.~Tanter, and C.~Demen{\'e},
  ``Adaptive spatiotemporal svd clutter filtering for ultrafast doppler imaging
  using similarity of spatial singular vectors,'' \emph{IEEE transactions on
  medical imaging}, vol.~37, no.~7, pp. 1574--1586, 2018.

\bibitem{loy2003fast}
G.~Loy and A.~Zelinsky, ``Fast radial symmetry for detecting points of
  interest,'' \emph{IEEE Transactions on pattern analysis and machine
  intelligence}, vol.~25, no.~8, pp. 959--973, 2003.

\bibitem{wang2004image}
Z.~Wang, A.~C. Bovik, H.~R. Sheikh, and E.~P. Simoncelli, ``Image quality
  assessment: from error visibility to structural similarity,'' \emph{IEEE
  transactions on image processing}, vol.~13, no.~4, pp. 600--612, 2004.

\bibitem{hahne2024learning}
C.~Hahne, ``Learning high-resolution delay-and-sum beamforming,'' in
  \emph{Bildverarbeitung f{\"u}r die Medizin 2024:
  Algorithmen-Systeme-Anwendungen. Proceedings des Workshops vom 10. bis 12.
  M{\"a}rz 2024 in Nürnberg}.\hskip 1em plus 0.5em minus 0.4em\relax Springer,
  2024, "in press", pp. 1--6.

\bibitem{rao2022ultrasonic}
J.~Rao, H.~Qiu, G.~Teng, R.~Al~Mukaddim, J.~Xue, and J.~He, ``Ultrasonic array
  imaging of highly attenuative materials with spatio-temporal singular value
  decomposition,'' \emph{Ultrasonics}, vol. 124, p. 106764, 2022.

\bibitem{Pialot:2023}
B.~Pialot, C.~Lachambre, A.~L. Mur, L.~Augeul, L.~Petrusca, A.~Basarab, and
  F.~Varray, ``Adaptive noise reduction for power doppler imaging using svd
  filtering in the channel domain and coherence weighting of pixels,''
  \emph{Physics in Medicine \& Biology}, vol.~68, no.~2, p. 025001, jan 2023.

\end{thebibliography}

\end{document}